\documentclass[a4paper,11pt]{article}
\pdfoutput=1

\usepackage{jheppub} 
\usepackage{graphicx,color}
\usepackage{enumitem}
\usepackage{braket}
\usepackage{slashed}
\usepackage[utf8]{inputenc}
\usepackage{hyperref}
\usepackage{float}
\usepackage{caption}
\usepackage{subfigure}
\usepackage{makecell}
\usepackage{epsfig}
\usepackage{amsmath,amssymb}
\usepackage{hyperref}
\hypersetup{
    colorlinks=flase,
    linkcolor=black,
    filecolor=magenta,      
    urlcolor=blue,
}

\hypersetup{
	colorlinks,
	citecolor=black,
	filecolor=black,
	linkcolor=blue,
	urlcolor=black
}

\newcommand{\bea}{\begin{eqnarray}}
\newcommand{\eea}{\end{eqnarray}}
\newcommand{\be}{\begin{equation}}
\newcommand{\ee}{\end{equation}}
\newcommand{\ba}{\begin{align}}
\newcommand{\ea}{\end{align}}
\newcommand{\ben}{\begin{enumerate}}
\newcommand{\een}{\end{enumerate}}
\newcommand{\bi}{\begin{itemize}}
\newcommand{\ei}{\end{itemize}}
\newcommand{\comments}[1]{}
\def\nn{\nonumber}
\def\pref#1{(\ref{#1})}
\def\hf{\frac12}

\def\bel#1{\begin{equation} \label{#1}}

\def\vo{\mathcal{V}}

\def\slz{SL(2, {\mathbb{Z}})}

\def\KKLT{{\scriptscriptstyle KKLT}}
\def\LVS{{\scriptscriptstyle LVS}}
\def\BG{{\scriptscriptstyle BG}}
\def\EW{{\scriptscriptstyle EW}}
\def\PERT{{\scriptscriptstyle PERT}}

\title{Moduli Stabilisation and the Statistics of SUSY Breaking in the Landscape}

\author[a]{Igor Broeckel,}
\author[a]{Michele Cicoli,}
\author[b]{Anshuman Maharana,}
\author[b]{Kajal Singh}
\author[c]{and Kuver Sinha}
\affiliation[a]{Dipartimento di Fisica e Astronomia, Universitá di Bologna, via Irnerio 46, 40126 Bologna, Italy and INFN, Sezione di Bologna, viale Berti Pichat 6/2, 40127 Bologna, Italy}
\affiliation[b]{Harish-Chandra Research Institute, HBNI, Jhunsi, Allahabad, UP 211019, India}
\affiliation[c]{Department of Physics and Astronomy, University of Oklahoma, Norman, OK 73019, USA}

\emailAdd{igor.broeckel@bo.infn.it}
\emailAdd{michele.cicoli@unibo.it}
\emailAdd{anshumanmaharana@hri.res.in}
\emailAdd{kajalsingh@hri.res.in}
\emailAdd{kuver.sinha@ou.edu}

\abstract{The statistics of the supersymmetry breaking scale in the string landscape has been extensively studied in the past finding either a power-law behaviour induced by uniform distributions of F-terms or a logarithmic distribution motivated by dynamical supersymmetry breaking. These studies focused mainly on type IIB flux compactifications but did not systematically incorporate the K\"ahler moduli. In this paper we point out that the inclusion of the K\"ahler moduli is crucial to understand the distribution of the supersymmetry breaking scale in the landscape since in general one obtains unstable vacua when the F-terms of the dilaton and the complex structure moduli are larger than the F-terms of the K\"ahler moduli. After taking K\"ahler moduli stabilisation into account, we find that the distribution of the gravitino mass and the soft terms is power-law only in KKLT and perturbatively stabilised vacua which therefore favour high scale supersymmetry. On the other hand, LVS vacua feature a logarithmic distribution of soft terms and thus a preference for lower scales of supersymmetry breaking. Whether the landscape of type IIB flux vacua predicts a logarithmic or power-law distribution of the supersymmetry breaking scale thus depends on the relative preponderance of LVS and KKLT vacua.}

\begin{document} 
\maketitle
\flushbottom

\section{Introduction}

For several decades, the idea of supersymmetry has been one of the central ideas in both phenomenological and formal aspects of high energy physics. From the point of view of phenomenology, it furnishes an elegant solution to the gauge hierarchy problem and provides natural dark matter candidates. Furthermore, the theory is supported by several sets of data via radiative corrections: gauge coupling unification, the value of the top mass, and the value of the Higgs mass which falls within the window allowed by the Minimal Supersymmetric Standard Model (MSSM). For a detailed discussion of the recent status of supersymmetric phenomenology, see \cite{Baer:2020kwz} and references therein. From a more formal point of view, supersymmetry plays a key r\^ole in making string theory a consistent theory of quantum gravity. (Approximately) supersymmetric string compactifications are typically stable, as supersymmetry protects solutions from various instabilities. Supersymmetric partners of the Standard Model (SM) are being actively searched for at the LHC, with null results thus far. Given this, the time is ripe to rethink the following  question: At what scale should we expect to find supersymmetry?

It is important to understand if string theory can provide guidance in this regard. The literature on supersymmetry breaking and its mediation in string theory is vast, much of it focused on constructions of specific supersymmetry breaking and MSSM-like sectors (see \cite{Maharana:2012tu, Denef:2007pq, Grana:2005jc, Silverstein:2004id, Frey:2003tf} for a review of these and other aspects of string phenomenology). A complementary line of inquiry, starting with the seminal work \cite{Bousso:2000xa, Feng:2000if, Ashok:2003gk, Douglas:2004qg, Douglas:2004zg, Denef:2004dm, Denef:2004ze, Denef:2004cf, Giryavets:2004zr, Misra:2004ky, Conlon:2004ds, Acharya:2005ez}, has been to frame the question in terms of statistical distributions in the landscape of flux vacua \cite{Douglas:2006es}. As described in \cite{Douglas:2004qg}, this program relies on several features of flux compactifications: they are the most well-understood string compactifications with moduli stabilisation and broken supersymmetry and thus provide a fertile arena where quantitative answers may be extracted; there are many vacua that at least roughly match the SM; the number of vacua is so large that statistical solutions make sense; and no single vacuum is favoured by the theory. These studies found a preference for high scale supersymmetry due to a uniform distribution of the supersymmetry breaking scale \cite{Douglas:2004qg, Denef:2004ze, Denef:2004cf}. This result has been obtained by taking the distribution of the relevant F-terms to be as given by the dilaton and complex structure F-terms, while the K\"ahler moduli F-terms have been neglected since these fields are stabilised only beyond tree-level.  

The purpose of this paper is to revisit the statistical distribution of the supersymmetry breaking scale in the type IIB flux landscape, paying particular attention to the stabilisation of the K\"ahler moduli. The motivation for our work comes from the fact that the dilaton and complex structure F-terms, if non-zero, typically give rise to a runaway for the K\"ahler moduli, unless they are tuned to be small as in a recent dS uplifting proposal \cite{Gallego:2017dvd}. This implies that stable vacua where moduli stabilisation is under control require the dilaton and complex structure F-terms to be suppressed with respect to the F-terms of the K\"ahler moduli. It is therefore the distribution of the F-terms of the K\"ahler moduli which determines the statistics of the supersymmetry breaking scale in the landscape. 

More precisely, in type IIB flux compactifications the complex structure moduli and the dilaton are fixed supersymmetrically at semi-classical level by 3-form fluxes \cite{Giddings:2001yu}. As we pointed out above, this supersymmetric stabilisation ensures the absence of instabilities along the K\"ahler moduli directions which are flat at tree-level due to the well-known `no-scale' property of the low-energy effective action \cite{Cremmer:1983bf, Ellis:1983sf, Burgess:1985zz, Burgess:2020qsc}. At this level of approximation, the cosmological constant vanishes and supersymmetry is broken due to non-zero F-terms of the K\"ahler moduli. However, due to the no-scale structure, the scale of the gravitino mass is unfixed and the soft terms might be zero (as in models where the SM is realised via D3-branes \cite{Marchesano:2004yn, Blumenhagen:2009gk,Conlon:2008wa, Aparicio:2014wxa}). The inclusion of no-scale breaking effects, which can come from either perturbative contributions to the K\"ahler potential or non-perturbative corrections to the superpotential, is therefore crucial to stabilise the K\"ahler moduli, to fix the supersymmetry breaking scale and to determine the soft terms. K\"ahler moduli stabilisation thus allows to write the gravitino mass (and consequently the soft terms) in terms of microscopic parameters like flux quanta or the number of D-branes. In turn, exploiting these relations and the knowledge of the distribution of these underlying parameters, one can deduce the distribution of the supersymmetry breaking scale in the landscape. 

We will try to perform a systematic study of the interplay between K\"ahler moduli stabilisation and the statistics of the supersymmetry breaking scale by considering three general scenarios: ($i$) models with purely non-perturbative stabilisation like in KKLT vacua \cite{Kachru:2003aw}; ($ii$) models where the K\"ahler moduli are frozen by balancing perturbative against non-perturbative effects as in the Large Volume Scenario (LVS) \cite{Balasubramanian:2005zx}; and ($iii$) models with purely perturbative stabilisation \cite{Berg:2005yu}. We primarily study the distributions focusing on vacua with zero cosmological constant, and do not explore the joint distribution of the cosmological and supersymmetry breaking scale in detail (although in the case of LVS we argue that the distribution of the supersymmetry breaking scale should remain the same for a wide range of values of the cosmological constant, see below).

Interestingly, we find that KKLT and perturbatively stabilised vacua behave similarly since in both cases the gravitino mass is governed by flux-dependent parameters (as the vacuum expectation value of the tree-level superpotential in KKLT models) which are uniformly distributed. Hence the statistics of supersymmetry breaking obeys a power-law behaviour implying that in these cases high scale supersymmetry is preferred, unless tempered by anthropics \cite{Baer:2017uvn}. Notice that these results match those derived in \cite{Douglas:2004qg} since in these cases the F-terms of the K\"ahler moduli, similarly to the dilaton and complex structure F-terms, turn out to be uniformly distributed. 

The situation in LVS models is instead different. In fact, we find that in this case the distribution of the supersymmetry breaking scale is exponentially sensitive to the distribution of the string coupling. Due to the exponential behaviour and the fact that the string coupling is uniformly distributed as a flux-dependent variable, the distribution of the soft terms turns out to be only logarithmic. This dependence gives rise to a large number of vacua with low-energy supersymmetry and reproduces in detail previous expectations following an intuition based on dynamical supersymmetry breaking \cite{Dine:2004is, Dine:2005iw, Dine:2005yq, Dine:2004ct} (although a significant difference is that \cite{Dine:2004is, Dine:2005iw} found a logarithmic distribution even in the case of KKLT, which we do not find).\footnote{We refer to \cite{Susskind:2004uv, Banks:2003es} for other early studies in this general direction.} 

LVS models are particularly interesting also because they provide examples where a crucial assumption formulated in \cite{Douglas:2004qg} can be explicitly shown to hold. This is the assumption that the distribution of the supersymmetry breaking scale is decoupled from the one of the cosmological constant. This was justified in \cite{Douglas:2004qg} by relying on the possible existence of several hidden sector models which contribute to the vacuum energy but not to supersymmetry breaking. In LVS models the depth of the non-supersymmetric AdS vacuum scales as $V_\LVS \sim - m_{3/2}^3 M_p$, where $m_{3/2}$ is the gravitino mass and $M_p$ the Planck scale. Hence any hidden sector responsible for achieving a nearly Minkowski vacuum contributes to the scalar potential with an F-term that scales as $F_{\rm hid}\sim m_{3/2}^{3/2} M_p^{1/2}$. In turn, in a typical gravity mediation scenario, the contribution to the soft terms from this hidden sector would be suppressed with respect to the gravitino mass since $M_{\rm soft}\sim F_{\rm hid}/M_p \sim \epsilon\, m_{3/2} \ll m_{3/2}$ with $\epsilon = \sqrt{m_{3/2}/M_p}\ll 1$.\footnote{An exception to this argument could however come from models where the SM is built via D3-branes at singularities which are sequestered from the sources of supersymmetry breaking in the bulk \cite{Blumenhagen:2009gk, Aparicio:2014wxa}.} Note that this implies that the distribution of the supersymmetry breaking scale is the same for all vacua with cosmological constant in the range $\pm V_\LVS$.

We have therefore shown that, while two alternative statistics of the supersymmetry breaking scale have been advanced before in the literature (power-law distributions by assuming democratic distributions of complex structure F-terms and logarithmic distributions by appealing to dynamical supersymmetry breaking), the different behaviours are neatly categorized by different stabilisation mechanisms. In order to determine if the distribution of the supersymmetry breaking scale is power-law or logarithmic, one should therefore determine the relative preponderance of LVS and KKLT vacua in the type IIB landscape. Given that LVS models do not rely on any tuning of the tree-level superpotential, one would naively expect them to arise much more frequently, so favouring a logarithmic distribution of the soft terms. However, a full understanding of this question requires detailed (numerical) studies of the distributions of flux vacua which is well beyond the scope of the present work. For estimates of the number of vacua as a function of the flux superpotential and the string coupling see \cite{DeWolfe:2004ns, Cicoli:2013cha, Demirtas:2019sip, Cole:2019enn}.

We finally point out that the ultimate goal of this line of research is to identify the mass scale of the supersymmetric particles preferred by the string landscape in order to find some guidance for low-energy searches of superpartners. In order to achieve this task, one has not just to understand the distribution of vacua, but has to focus also on \textit{phenomenologically viable} vacua. This means that one should impose additional constraints coming for example from cosmology or from anthropic arguments \cite{Baer:2017uvn}. For example, in string compactifications both the moduli masses and the soft terms turn out to be of order the gravitino mass. Hence the absence of any cosmological moduli problem \cite{Coughlan:1983ci, Banks:1993en, deCarlos:1993wie, Kane:2015jia}, which requires moduli masses above $\mathcal{O}(50)$ TeV, tends to push the soft terms considerably above the TeV-scale unless the SM sector is sequestered from supersymmetry breaking (as in some D3-brane models \cite{Blumenhagen:2009gk, Aparicio:2014wxa}.) We leave a detailed study of these additional phenomenological and cosmological constraints for future work. 

This paper is organised as follows. In Sec. \ref{Sec2} we first review previous determinations of the statistics of the supersymmetry breaking scale neglecting the K\"ahler moduli. After explaining why this analysis is incomplete and a more accurate study should take the K\"ahler moduli into account, we then provide an overview of the three general classes of K\"ahler moduli stabilisation schemes mentioned above: KKLT \cite{Kachru:2003aw}, LVS \cite{Balasubramanian:2005zx} and perturbative stabilisation \cite{Berg:2005yu}. In Sec. \ref{Sec3} we derive in detail the distribution of the supersymmetry breaking scale for each of these three scenarios, while in Sec. \ref{Sec4} we discuss the interplay between our results and previous findings in the literature and the implications of our distributions for phenomenology. Our conclusions are presented in Sec. \ref{Conclusions}. Finally App. \ref{gsdistribution} presents a discussion of the distribution of the string coupling while App. \ref{AppB} summarises the structure of the soft terms in KKLT and LVS models with an MSSM-like sector on either D3 or D7-branes.

\section{The importance of the K\"ahler moduli for the SUSY breaking statistics}
\label{Sec2}

The statistics of the supersymmetry breaking scale in the landscape has been investigated mainly in the context of type IIB flux compactifications since this is one of the best examples where moduli stabilisation can be achieved with control over the effective field theory. However previous studies focused only on the contribution to supersymmetry breaking from the axio-dilaton and the complex structure moduli, ignoring the dynamics of the K\"ahler moduli \cite{Ashok:2003gk, Douglas:2004qg, Douglas:2004zg, Denef:2004dm, Denef:2004ze, Denef:2004cf}. In what follows we shall instead point out that the K\"ahler moduli play a crucial r\^ole in determining the correct statistics of the supersymmetry breaking scale in the landscape. 

\subsection{SUSY breaking statistics neglecting the K\"ahler moduli}
\label{SUSYstatRev}

The starting point of our discussion is type IIB string theory compactified on a Calabi-Yau $X$ which, together with an appropriate orientifold involution, can lead to an $N=1$ supergravity effective action in 4D. One of the nicest features of these compactifications is that one can turn on RR and NSNS 3-form fluxes $F_3$ and $H_3$ without destroying the underlying Calabi-Yau structure since the flux backreaction just introduces warping \cite{Giddings:2001yu}. Moreover, these background 3-form fluxes, which appear in the combination $G_3 = F_3-iSH_3$, can stabilise the axio-dilaton $S$ and all complex structure moduli $U^{\alpha}$ (with $\alpha=1,...,h^{1,2}(X)$) by generating the following tree-level superpotential \cite{Gukov:1999ya}:
\be
W_{\rm tree} = \int_X G_3 \wedge \Omega(U)\,,
\label{Wtree}
\ee
where $\Omega(U^\alpha)$ is the holomorphic $(3,0)$-form of the Calabi-Yau $X$ that depends on the $U$-moduli.

The tree-level K\"ahler potential which can be obtained from direct dimensional reduction is instead \cite{Grimm:2004uq}:
\be
K_{\rm tree} = -2\ln\vo-\ln\left(S+\bar{S} \right)-\ln\left(-{\rm i}\int_X \Omega(U) \wedge \bar{\Omega}(\bar{U})\right)\,,
\label{Ktree}
\ee
where $\vo$ is the dimensionless volume of the internal manifold expressed in units of the string length $\ell_s=2\pi\sqrt{\alpha'}=M_s^{-1}$. The Calabi-Yau volume $\vo$ is also a function of the real parts of the K\"ahler moduli $T_i = \tau_i+{\rm i}\theta_i$ (with $i=1,...,h^{1,1}(X)$) where the $\tau_i$'s control the size of internal divisors while the $\theta_i$'s are the axions obtained from the dimensional reduction of the RR 4-form $C_4$ over the same 4-cycles. For the simplest cases with just a single K\"ahler modulus, $\vo = \tau^{3/2}$. 

The scalar potential is obtained by plugging the expressions (\ref{Wtree}) and (\ref{Ktree}) in the general expression of the F-term scalar potential in supergravity (setting $M_p\equiv 1/\sqrt{8\pi\,G_N} = 1$ and neglecting possible contributions coming from D-terms):
\be
V_F = e^K \left(K^{i\bar{j}}D_iW D_{\bar{j}}\overline{W}-3|W|^2 \right) = K_{i\bar{j}} F^i \overline{F}^{\bar{j}} - 3 m_{3/2}^2\,,
\label{VF}
\ee
where:
\be
F^i = e^{K/2}\,K^{i\bar{j}} D_{\bar{j}}\overline{W}\qquad\text{and}\qquad m_{3/2} = e^{K/2}|W|\,.
\label{m32}
\ee
Given that the tree-level K\"ahler potential (\ref{Ktree}) factorises, the F-term scalar potential (\ref{VF}) takes the form (denoting all complex structure and K\"ahler moduli collectively as $U$ and $T$ respectively):
\be
V_{\rm tree} = |F^S|^2 + |F^U|^2 + |F^T|^2 - 3 m_{3/2}^2\,.
\label{VFtree}
\ee
Ref. \cite{Douglas:2004qg, Denef:2004ze, Denef:2004cf} considered situations where supersymmetry is spontaneously broken at the minima of the scalar potential (\ref{VFtree}) and studied the distribution of the supersymmetry breaking scale taking the distribution of the relevant F-terms to be that  obtained from the analysis for the $S$ and $U$-moduli. The K\"ahler moduli have been instead neglected since these moduli are not stabilised by fluxes at tree-level, and so the dynamics that fixes them beyond the tree-level approximation has been assumed to give rise just to small corrections to the leading order picture. 

Hence the distribution of supersymmetry breaking vacua has been claimed to be given by \cite{Douglas:2004qg}:
\be 
dN (F ,\hat\Lambda) = \prod d^2F^S\, d^2 F^U\, d\hat\Lambda \,\rho(F,\hat\Lambda)\,,
\label{dd1}
\ee
where $\hat\Lambda$ is the depth of the supersymmetric AdS vacuum, $\hat{\Lambda} = 3 m_{3/2}^2$, and the F-terms of the $T$-moduli have been ignored. Requiring in addition a vanishing cosmological constant, one obtains:
\be
dN_{\Lambda=0}(F)  = \prod d^2F^S\, d^2 F^U\,d\hat\Lambda \,\rho(F,\hat\Lambda) \, \delta\left(|F^S|^2 + |F^U|^2 -\hat \Lambda\right).
\label{dd2}
\ee

Ref. \cite{Douglas:2004qg} makes two claims about the cosmological constant: the first claim is that the distribution of values of the supersymmetric AdS vacuum $\hat\Lambda  = - \Lambda = e^K\,|W|^2$ is determined by the distribution of the tree-level superpotential (\ref{Wtree}) which is uniformly distributed as a complex variable near zero, and throughout its range is more or less uniform. The second claim is instead that this distribution is relatively uncorrelated with the supersymmetry breaking parameters if the hidden sector which breaks supersymmetry is different from the one which is responsible to obtain a nearly zero cosmological constant. 

If one assumes a decoupling of the cosmological constant problem from the question of supersymmetry breaking, then the density function $\rho$ is in fact independent of $\hat\Lambda$, leading to:
\be
dN_{\Lambda=0}(F) =  d^2F \, \rho(F)\,,
\label{dd3}
\ee
where we have collectively denoted all the F-terms of the axio-dilaton and the complex structure moduli simply as $F$. Using the vanishing cosmological constant condition $|F|^2 = 3 m_{3/2}^2$ and the fact that $d^2 F \simeq |F|\,d|F| \simeq m_{3/2}\, d m_{3/2}$, (\ref{dd3}) reduces to:
\be
dN_{\Lambda=0}(m_{3/2}) \simeq  \rho(m_{3/2})\,m_{3/2}\, d m_{3/2}\,.
\label{dd3new}
\ee
Given that the gravitino mass is set by the F-terms of the axion-dilaton and the complex structure moduli, and $F^S$ and $F^U$ in type IIB flux vacua turn out to be uniformly distributed as complex variables, \cite{Douglas:2004qg} considered $\rho(m_{3/2})$ as independent on $m_{3/2}$. In order to keep this discussion more general in view of our results in the case where the $T$-moduli are included, we consider instead:
\be 
\rho(m_{3/2}) \, \sim \, m_{3/2}^\beta \qquad \text{with}\qquad\beta \geq 0\,,
\label{rhoDistr}
\ee
which implies:
\be
dN_{\Lambda=0}(m_{3/2}) \simeq m_{3/2}^{\beta + 1} \, d m_{3/2} \qquad \text{with}\qquad\beta \geq 0\,,
\label{GravDistr}
\ee
where $\beta=0$ for the case where the dynamics of the K\"ahler moduli is neglected \cite{Douglas:2004qg, Denef:2004ze, Denef:2004cf}. Notice that the result with $\beta=0$ would indicate a preference for high scale supersymmetry.

\subsection{SUSY breaking statistics including the K\"ahler moduli}

The importance of the K\"ahler moduli for the statistics of the supersymmetry breaking scale in the landscape can be easily understood by noticing that the tree-level superpotential (\ref{Wtree}) is independent on the $T$-moduli due to holomorphy combined with the axionic shift symmetry. Hence the F-terms of the K\"ahler moduli become $F^T = e^{K/2}\,\overline{W} K^{T\bar{T}} K_{\bar{T}}$ and the scalar potential (\ref{VFtree}) can be rewritten as:
\be
V_{\rm tree} = |F^S|^2 + |F^U|^2 + m_{3/2}^2 \left(K_{\bar{T}} K^{\bar{T}T} K_T - 3 \right) .
\label{VFtreenew}
\ee
A generic property of type IIB vacua which holds for all Calabi-Yau manifolds is the famous `no-scale' relation $K_{\bar{T}} K^{\bar{T}T} K_T=3$ which has been recently shown to be a low-energy consequence of the axionic shift symmetry combined with approximate higher dimensional symmetries like scale invariance and supersymmetry \cite{Burgess:2020qsc}. This no-scale property of type IIB vacua has important consequences which we now briefly discuss:
\bi
\item At tree-level the scalar potential (\ref{VFtreenew}) reduces to (where $K_{\rm cs}$ denotes the K\"ahler potential for the $U$-moduli):
\be
V_{\rm tree} = |F^S|^2 + |F^U|^2  = \frac{e^{K_{\rm cs}}}{\vo^2 \left(S+\bar{S}\right)}\left[|D_S W|^2+ |D_U W|^2 \right].
\label{VFtreenew2}
\ee
This result shows that any vacuum where either $D_S W\neq 0$ or $D_U W \neq 0$ is unstable since it gives rise to a run-away for the volume mode $\vo$ at tree-level. One could envisage a scenario where this run-away is counter-balanced by quantum corrections but when the perturbative expansion is under control these effects are expected to be subdominant by consistency. Hence a stable solution requires $F^S=F^U=0$.\footnote{See however \cite{Gallego:2017dvd} for dS uplifting models where $F^S$ and $F^U$ are tuned to very small values. These cases are consistent with our claims since they feature $F^S\sim F^U \ll F^T$.} 
This implies that the statistic of the supersymmetry breaking scale in the landscape should instead be driven by the F-terms of the K\"ahler moduli.

\item At tree-level, the gravitino mass is set by the F-terms of the $T$-moduli since the no-scale relation implies $|F^T|^2 = 3 m_{3/2}^2$. This is contrast with the case where the K\"ahler moduli are ignored and $m_{3/2}$ is set by the F-terms of $S$ and $U$-moduli. Thus there is no reason to expect that coefficient $\beta$ in the distribution of the gravitino mass (\ref{rhoDistr}) should be zero. Moreover, the K\"ahler moduli are still flat at tree-level, and so any scale of supersymmetry breaking is equally valid. To set $m_{3/2}$ and to understand its distribution one has therefore to study which corrections to the tree-level action can stabilise the K\"ahler moduli. We shall show that in a large number of flux vacua (all the LVS examples) $F^T$ is not uniformly distributed, and so $\beta \neq 0$. 

\item The gravitino mass does not necessarily fix the scale of the soft supersymmetry breaking terms in the visible sector. In fact, in type IIB models an MSSM-like visible sector can be located on either stacks of D7-branes with non-zero gauge fluxes or on D3-branes at singularities. The tree-level K\"ahler potential including D7 and D3 matter fields, respectively denoted as $\phi_3$ and $\phi_7$, is given by (focusing for simplicity on the case with $h^{1,1}(X)=1$) \cite{Jockers:2005pn}:
\be
K_{\rm tree} = -3 \ln\left(T+\bar{T} - \bar{\phi}_3 \phi_3 \right) - \ln\left(S + \bar{S} - \bar{\phi}_7 \phi_7 \right) 
\simeq K_0 + \tilde{K}_3\,\bar{\phi}_3 \phi_3 + \tilde{K}_7\, \bar{\phi}_7 \phi_7\,, \nn
\ee
where $K_0$ denotes the K\"ahler potential for $T$ and $S$ while $\tilde{K}_3 = 3 \left(T+\bar{T}\right)^{-1}$ and $\tilde{K}_7 = \left(S + \bar{S}\right)^{-1}$. On the other hand the visible sector gauge kinetic functions for D7s and D3s at tree-level read:
\be
f_3 = S\qquad\text{and}\qquad f_7 = T\,.
\ee
Moreover the general expressions of the soft scalar and gaugino masses in gravity mediation look like:
\be
m_0^2 = m^2_{3/2}-\overline{F}^{\bar{i}}F^j  \partial_{\bar{i}}\partial_j \ln \tilde{K}\qquad\text{and}\qquad 
M_{1/2} = \frac{1}{2\,{\rm Re}(f)}\,F^i\partial_i f\,.
\ee
Using $F^S=0$ and $F^T = e^{K/2}\,\overline{W} K^{T\bar{T}} K_{\bar{T}}$, we then end up with:
\bea
{\rm D3}:\quad m_0 &=& M_{1/2} = 0 \nn \\
{\rm D7}:\quad m_0 &=& |M_{1/2}| = m_{3/2} \,.
\eea
Hence we can clearly see that the soft terms are set by the gravitino mass only for D7s, while for D3s they are suppressed with respect to $m_{3/2}$. We conclude that the inclusion of perturbative and/or non-perturbative corrections to the 4D effective action which break the no-scale structure is crucial for two important tasks: ($i$) to stabilise the K\"ahler moduli, which in turn fixes the leading order value of $F^T$ and $m_{3/2}$; ($ii$) to generate a subleading shift to the tree-level results for $F^S$ and $F^T$ which yield non-zero contributions to $m_0$ and $M_{1/2}$ for visible sector models on D3-branes.
\ei

\subsection{Overview of type IIB K\"ahler moduli stabilisation}

After having motivated the importance of K\"ahler moduli stabilisation for understanding the correct distribution of the supersymmetry breaking scale in the type IIB flux landscape, we describe now the main features of three different classes of stabilisation scenarios classified in terms of perturbative and non-perturbative corrections to the 4D low-energy action.

\subsubsection{Purely non-perturbative stabilisation: KKLT}
 
Let us start by reviewing the KKLT \cite{Kachru:2003aw} stabilisation mechanism and identify the relevant parameters. The starting point is to introduce 3-form fluxes which stabilise the axio-dilaton and all complex structure moduli at $F^S=F^U=0$ \cite{Giddings:2001yu}. The next step is to allow for effects like gaugino condensation on D7 branes or Euclidean D3 instantons, both wrapped on internal 4-cycles. Both of these effects lead to non-perturbative corrections to the superpotential that stabilise the K\"ahler modulus $T = \tau+i\theta$ if the vacuum expectation value of $W_{\rm tree}$ is tuned to exponentially small values. Thus in KKLT models the K\"ahler potential takes the tree-level expression given in (\ref{Ktree}) while the superpotential is:
\be
W = W_0 + A\,e^{-a\,T}\,,
\label{WKKLT}
\ee
where $W_0$ is the vacuum expectation value of the tree-level superpotential (\ref{Wtree}). Moreover $a= 2\pi/\mathfrak{n}$ with $\mathfrak{n}=1$ for stringy instantons while in the case of more standard field theoretic non-perturbative effects on stacks of D7-branes $\mathfrak{n}$ is related to the number of D7-branes that, together with the orientifold involution, determines the rank of the condensing gauge group (for example for gaugino condensation in a pure $SU(N)$ super Yang-Mills theory $\mathfrak{n}=N$). The scalar potential is obtained by plugging the expressions (\ref{Ktree}) and (\ref{WKKLT}) in the general expression of the F-term supergravity scalar potential (\ref{VF}).  After minimising with respect to the axion $\theta$, one arrives at (with $s = {\rm Re}(S)$):
\be
V_\KKLT = \frac{2e^{-2a\tau}a^2A^2}{3s\vo^{2/3}}\left(1 +\frac{3}{a \tau}\right)-\frac{2e^{-a\tau}aAW_0}{s\vo^{4/3}}\,,
\ee
where $\vo=\tau^{3/2}$ is the dimensionless CY volume in units of the string length $\ell_s=2\pi\sqrt{\alpha'}=M_s^{-1}$. Minimising this potential with respect to the volume we get the relation:
\be
e^{a\langle\tau\rangle} = \frac{2 A a \langle\tau\rangle}{3 W_0}\left(1 + \frac{3}{2 a \langle\tau\rangle}\right) \simeq \frac{2 A a \langle\tau\rangle}{3 W_0}\qquad\Leftrightarrow\qquad \langle\tau\rangle \simeq \frac{1}{a}\,|\ln W_0|\,,
\label{KKLTmin}
\ee 
where we took the limit $a\langle\tau\rangle \gg 1$ where higher instantons corrections to (\ref{WKKLT}) can be safely ignored and we considered natural values of the prefactor $A$ of the non-perturbative contribution to $W$, i.e. $A\sim\mathcal{O}(1)$. Notice that (\ref{KKLTmin}) leads to two important observations:
\ben
\item A minimum at values of $\langle\tau\rangle \gg 1$, where stringy corrections to the effective action can be neglected, can be obtained only if $W_0$ is tuned to exponentially small values. Notice that such a tuning guarantees also the consistency of  neglecting perturbative corrections to $K$ (since they give rise to contribution to $V$ which are proportional to $|W_0|^2$).

\item This vacuum preserves supersymmetry since (\ref{KKLTmin}) implies $F^T=0$. Hence, as can be seen from (\ref{VF}), the vacuum energy is negative with $V = -3\, m_{3/2}^2$ where in this case $m_{3/2}$ should just be intended as the parameter defined in (\ref{m32}) without any reference to the gravitino mass.
\een
A Minkowski or slightly dS vacuum can be obtained by adding to the scalar potential the positive definite contribution coming from D3-branes at the end of a warped throat \cite{Kachru:2003aw} (another interesting option relies on $\alpha'$ corrections to $K$ \cite{Westphal:2006tn}). As shown in \cite{Aparicio:2015psl}, this requires the addition of a nilpotent superfield in the 4D effective field theory description. The presence of this nilpotent superfield gives rise to a Minkowski vacuum where the relation (\ref{KKLTmin}) gets modified to:
\be
e^{a\langle\tau\rangle} = \frac{2 A a \langle\tau\rangle}{3 W_0}\left(1 + \frac{5}{2 a \langle\tau\rangle}\right).
\label{KKLTmindS}
\ee 
Interestingly, (\ref{KKLTmin}) and (\ref{KKLTmindS}) agree at leading order, and so we can safely consider $\langle\tau\rangle \simeq \frac{1}{a}\,|\ln W_0|$ also at the Minkowski minimum where supersymmetry is broken. In this case the gravitino mass becomes (where the vacuum expectation value of $s$ sets the string coupling, i.e. $s = g_s^{-1}$):
\be
m_{3/2} \simeq \sqrt{\frac{g_s}{8\pi}}\frac{|W_0|}{\langle\vo\rangle} \simeq \frac{\pi\,g_s^{1/2}}{\mathfrak{n}^{3/2}}\frac{|W_0|}{|\ln W_0|^{3/2}}\,.
\label{m32KKLT}
\ee
This equation shows clearly that, begin exponentially small, it is $W_0$ that determines the order of magnitude of $m_{3/2}$. The soft terms in the KKLT scenario can be generated via either gravity or anomaly mediation \cite{Aparicio:2015psl, Choi:2005ge} with the MSSM-like visible sector located on either stacks of D7-branes with non-zero gauge fluxes or on D3-branes at singularities. In both cases, the overall scale of the soft terms $M_{\rm soft}$ is of order the gravitino mass up to a possible 1-loop factor whose presence is model-dependent: $M_{\rm soft}\sim m_{3/2}$.

\subsubsection{Perturbative vs non-perturbative effects: LVS}
 
The starting point of LVS models is the same as in KKLT constructions since at tree-level the complex structure moduli and the dilaton are stabilised supersymmetrically by non-zero 3-form fluxes at $F^U=0$ and $F^S=0$. At this semi-classical level of approximation, the K\"ahler moduli are however flat directions due to the underlying no-scale cancellation which is inherited from higher-dimensional rescaling symmetries \cite{Burgess:2020qsc}. 

The simplest LVS model (see \cite{Cicoli:2008va, Cicoli:2011yy, Cicoli:2016chb, AbdusSalam:2020ywo} for more general constructions) features 2 K\"ahler moduli and a CY volume of the form $\vo=\tau_b^{3/2}-\tau_s^{3/2}$ where $\tau_b$ is a `big' divisor controlling the overall volume while $\tau_s$ is a `small' divisor supporting non-perturbative effects, with $\tau_b\gg \tau_s\gg 1$ \cite{Balasubramanian:2005zx}. If the leading order $\alpha'$ correction to the effective action is included, the K\"ahler and superpotential of LVS models look like:
\bea
K &=& -2\ln\left(\vo+\frac{\xi}{2}\left(\frac{S+\bar{S}}{2}\right)^{3/2}\right)-\ln\left(S+\bar{S} \right)-\ln\left(-i\int_X \Omega(U) \wedge \bar{\Omega}(\bar{U})\right)\\
W &=& W_0+A_s\,e^{-a_sT_s}\,,
\eea
with $a_s= 2\pi/\mathfrak{n}$ as in the KKLT case and $\xi\equiv -\frac{\chi(X) \zeta(3)}{2(2\pi)^3}$ where $\chi(X)$ is the CY Euler number and $\zeta$ is the Riemann zeta function. Notice that $A_s$ and $\xi$ are both expected to be $\mathcal{O}(1)$ parameters. After setting $S$ and all the $U$-moduli at their flux-stabilised values and fixing the axionic partner of $\tau_s$ at its minimum, the scalar potential (\ref{VF}) takes the form:
\be
V_\LVS=\frac43\frac{a_s^2A_s^2\sqrt{\tau_s}e^{-2a_s\tau_s}}{s\mathcal{V}}-\frac{2a_sA_s|W_0|\tau_se^{-a_s\tau_s}}{s\mathcal{V}^2}+\frac{3\sqrt{s}\xi|W_0|^2}{8\vo^3}\,.
\label{VLVS}
\ee
Minimising the potential we obtain the following conditions on the moduli (with $s=g_s^{-1}$):
\be
\langle\vo\rangle \simeq \frac{3\sqrt{\langle\tau_s\rangle}\,|W_0|}{4a_s A_s}\,e^{a_s \langle\tau_s\rangle}\qquad\text{and}\qquad \langle\tau_s\rangle \simeq \frac{1}{g_s}\left(\frac{\xi}{2}\right)^{2/3}\,.
\label{LVSmin}
\ee
Let us again stress two important points which follow from (\ref{LVSmin}):
\ben
\item In LVS models, it is the smallness of $g_s$ that guarantees that the effective field theory is under control. In fact, if the string coupling is such that perturbation theory does not break down, i.e. $g_s\lesssim 0.1$, stringy corrections to the 4D action can be safely ignored since both $\tau_b$ and $\tau_s$ are much larger than the string scale. Hence these models can exist for natural values of the flux-generated superpotential $W_0$ with $W_0 \sim \mathcal{O}(1-10)$. 

\item The LVS vacuum is AdS with $V_{\rm LVS}\sim - m_{3/2}^3$ and non-supersymmetric with the largest F-term given by $F^{T_b} \sim \tau_b m_{3/2}$. Hence the Goldstino is the fermionic partner of $T_b$ in the corresponding $N=1$ chiral superfield. This is eaten up by the gravitino which acquires a non-zero mass.
\een
As in KKLT models, an additional positive definite contribution to the scalar potential has to be added in order to obtain a Minkowski solution. Several `uplifting' mechanisms have been proposed and the main ones involve anti-branes \cite{Kachru:2003aw}, T-branes \cite{Cicoli:2015ylx}, hidden sector non-perturbative effects \cite{Cicoli:2012fh} or non-zero F-terms of the dilaton and complex structure moduli \cite{Gallego:2017dvd}. The important observation here is that all these mechanisms modify the relations in (\ref{LVSmin}) only at subleading order. Hence we can consider (\ref{LVSmin}) a good analytic estimate also for the location of the Minkowski minimum. Thus the gravitino mass becomes:
\be
m_{3/2} \simeq \sqrt{\frac{g_s}{8\pi}}\frac{|W_0|}{\langle\vo\rangle} \simeq c_1\,\frac{g_s}{\mathfrak{n}}\,e^{- \frac{c_2}{g_s\mathfrak{n}}}\,,
\label{m32LVS}
\ee
where $c_1$ and $c_2$ are $\mathcal{O}(1)$ parameters given by:
\be
c_1 = \frac{\sqrt{8\pi} A_s}{3}\,\left(\frac{2}{\xi}\right)^{1/3}\qquad\text{and}\qquad c_2 = 2\pi \left(\frac{\xi}{2}\right)^{2/3} \,.
\ee
Contrary to KKLT scenarios where the value of $m_{3/2}$ was determined by $W_0$, (\ref{m32LVS}) shows clearly that in LVS models the scale of the gravitino mass is set by the string coupling. Another difference between KKLT and LVS models, is that in LVS constructions the contribution to the soft terms from anomaly mediation is always loop-suppressed with respect to the contribution from gravity mediation (since similar cancellations in both mediation mechanisms take place due to the underlying no-scale property of these vacua). Moreover, in LVS models, the overall scale of the soft terms depends crucially on the fact that the SM is realised on either D7 or D3-branes \cite{Blumenhagen:2009gk, Aparicio:2014wxa, Conlon:2005ki, Conlon:2006wz}:
\be
{\rm D7}:\,\,M_{\rm soft}\sim m_{3/2}\qquad \qquad{\rm D3}:\,\,M_{1/2}\sim m_{3/2}^2 \quad\text{and}\quad m_0 \sim m_{3/2}^p\,,
\label{SofttermsLVS}
\ee
where $p$ can be either $p=2$ or $p=3/2$ depending on the mechanism considered to obtain a Minkowski vacuum \cite{Aparicio:2014wxa}.

\subsubsection{Purely perturbative stabilisation: $\alpha'$ vs $g_s$ effects}
\label{PertSec}
 
Let us now describe K\"ahler moduli stabilisation based just on perturbative corrections to the effective action \cite{Berg:2005yu}. As shown in \cite{Conlon:2005ki}, when $W_0$ takes natural $\mathcal{O}(1-10)$ values and no blow-up modes like the `small' modulus $\tau_s$ of LVS models are present, non-perturbative effects are subdominant with respect to perturbative corrections in either $\alpha'$ or $g_s$. 

The main perturbative corrections to $K$ which yield non-zero contributions to the scalar potential are (for an more detailed discussion of these effects see \cite{Cicoli:2018kdo, Burgess:2020qsc}): $\mathcal{O}(\alpha'^3)$ corrections at tree-level in $g_s$ computed in \cite{Becker:2002nn} and open string 1-loop effects at both $\mathcal{O}(\alpha'^2)$ and $\mathcal{O}(\alpha'^4)$ computed in \cite{Berg:2005ja}. In the simplest case of a single K\"ahler modulus, these corrections to $K$ take the form \cite{Becker:2002nn, Berg:2005ja, Berg:2007wt}:
\be
K_{g_s^0\alpha'^3} = -\frac{\xi}{g_s^{3/2} \vo}\,,\qquad K_{g_s^2\alpha'^2} = g_s \,\frac{b(U)}{\vo^{2/3}}\,,\qquad 
K_{g_s^2\alpha'^4} = \frac{c(U)}{\vo^{4/3}}\,.
\ee
The parameters $b(U)$ and $c(U)$ are in general unknown functions of the complex structure moduli (and open string moduli as well) which have been computed explicitly only for simple toroidal orientifolds like $\mathbb{T}^6 / (\mathbb{Z}_2 \times \mathbb{Z}_2)$ \cite{Berg:2005ja}. They are however expected to be $\mathcal{O}(1-10)$ numbers in absence of fine tuning. Interestingly, the $\mathcal{O}(g_s^2\alpha'^2)$ corrections to $K$ proportional to $b(U)$ experience an `extended no-scale' cancellation \cite{Cicoli:2007xp}, and so they contribute to the scalar potential only at $\mathcal{O}(g_s^4\alpha'^4)$. Hence we can neglect them since for $g_s \lesssim 0.1$ they are subleading with respect to the correction to $K$ proportional to $c(U)$. 

After minimising the scalar potential with respect to the axio-dilaton and the complex structure moduli by solving $D_S W=D_U W=0$, the potential for the K\"ahler modulus is given by:
\be
V = g_s\,\frac{|W_0|^2}{\vo^3}\left(-\frac{3|\xi|}{8 g_s^{3/2}} + \frac{c(U)}{\vo^{1/3}}\right),
\label{Vpert}
\ee
where we have considered a negative value of the coefficient $\xi$ in order to get a minimum.\footnote{Notice that $\xi<0$ would require $h^{1,2}<h^{1,1}$ which for $h^{1,1}=1$ would work only for rigid CY manifolds without complex structure moduli, i.e. for $h^{1,2}=0$. However the potential (\ref{Vpert}) could also describe a more general situation with $h^{1,1}\gg 1$ where all K\"ahler moduli scale in the same way, i.e. $\tau_i \sim \vo^{2/3}$ $\forall i=1,...,h^{1,1}$.} Minimising with respect to $\vo$ we obtain a non-supersymmetric (since $F^T\neq 0$) AdS vacuum at:
\be
\langle \vo \rangle \simeq 26\,g_s^{9/2}\,\left(\frac{c}{|\xi|}\right)^3\,.
\label{PertMin}
\ee
Let us make again two important considerations:
\ben
\item The parameter controlling the string loop expansion is $g_s$ while the $\alpha'$ expansion is controlled by $\vo^{-1/3}$. Hence perturbation theory does not break down if $g_s \ll 1$ and $\vo \gg 1$. The first of these two conditions can be satisfied by an appropriate choice of 3-form fluxes which stabilise ${\rm Re}(S)=g_s^{-1}$. On the other hand, the second condition, as can be seen in (\ref{PertMin}), requires the parameter $c$ to be tuned such that $c \sim g_s^{-(3/2+q)}\gg 1$ with $q>0$ (for $|\xi|\sim \mathcal{O}(1)$). In fact, plugging this relation in (\ref{PertMin}) one obtains $\langle \vo \rangle \simeq 26\,g_s^{-3 q}\gg 1$ for $g_s \ll 1$. Given that $c=c(U)$ is a function of the complex structure moduli which are fixed in terms of flux quanta, we expect this tuning to be possible in the string landscape by scanning through different combinations of flux quanta. 

\item The minimum in (\ref{PertMin}) is non-supersymmetric, since $F^T\neq 0$, and AdS since $\langle V\rangle \simeq - 0.1\,c\,g_s\,|W_0|^2\,\langle\vo\rangle^{-10/3}$.
\een
The vacuum energy can be set to zero via the same uplifting mechanisms mentioned for KKLT and LVS models which are expected to yield only subleading corrections to the location of the minimum in (\ref{PertMin}). Hence the gravitino mass turns out to be:
\be
m_{3/2} \simeq \sqrt{\frac{g_s}{8\pi}}\frac{|W_0|}{\langle\vo\rangle} \simeq \lambda\,\frac{|W_0|}{g_s^4\,c^3}\qquad \text{with}\qquad \lambda \sim \mathcal{O}(10^{-2}) \,.
\label{m32Pert}
\ee
In this case it is the tuned parameter $c$ which controls the order of magnitude of the gravitino mass. The generation of the soft terms in these models with purely perturbative stabilisation of the K\"ahler moduli has not been studied. However we expect them to have the same behaviour as in (\ref{SofttermsLVS}) for LVS models since the contribution from anomaly mediation should feature a leading order cancellation due to the no-scale structure also in this case where therefore the soft terms are generated from gravity mediation.

\section{SUSY breaking statistics with K\"ahler moduli stabilisation} 
\label{Sec3}

In Sec. \ref{Sec2} we have first explained why a proper understanding of the statistics of the supersymmetry breaking scale in the type IIB flux landscape necessarily requires the inclusion of the K\"ahler moduli, and we have then illustrated the key-features of the main K\"ahler moduli stabilisation mechanisms based on different combinations of perturbative and non-perturbative corrections to the 4D effective field theory. In this section we shall instead determine the actual distribution of the gravitino mass, i.e. the actual value of the coefficient $\beta$ in (\ref{rhoDistr}), for each of these scenarios separately.

\subsection{LVS models}
\label{LVSsection}

Let us start our analysis of the distribution of the gravitino mass by focusing first on LVS models since they do not require any tuning of the tree-level flux superpotential. In these scenarios the minimum and $m_{3/2}$ are given respectively by (\ref{LVSmin}) and (\ref{m32LVS}). Notice that $m_{3/2}$ in (\ref{m32LVS}) does not depend on $|W_0|$ contrary to the expression (\ref{m32KKLT}) of the gravitino mass in KKLT models which is mainly determined by $|W_0|$. 

Varying the gravitino mass with respect to the flux-dependent parameter $g_s$ and the integer parameter $\mathfrak{n}$ which encodes the nature of non-perturbative effects, and working in the limit $a_s \tau_s \gg 1$ where the instanton expansion is under control, i.e. for $c_2 \gg g_s \mathfrak{n}$, we obtain:
\bea
d m_{3/2} &=& \frac{\partial m_{3/2}}{\partial g_s}\,d g_s + \frac{\partial m_{3/2}}{\partial \mathfrak{n}}\, d \mathfrak{n}
\simeq c_2\, \frac{m_{3/2}}{(g_s \mathfrak{n})^2} \left(\mathfrak{n}\,d g_s + g_s \,d \mathfrak{n} \right) \nn \\
&\simeq& m_{3/2} \left[\ln\left(\frac{M_p}{m_{3/2}}\right)\right]^2 \left(\mathfrak{n}\,d g_s + g_s \,d \mathfrak{n} \right),
\label{m32distr}
\eea
where in the last step we have introduced Planck units and we have approximated $m_{3/2} \sim M_p\,e^{- \frac{c_2}{g_s\mathfrak{n}}}$.

As we discuss in App. \ref{gsdistribution}, the distribution of the string coupling can be considered as approximately uniform\footnote{
In App. \ref{gsdistribution}, we numerically study this distribution for rigid Calabi-Yaus and find a uniform distribution. The analysis
for general Calabi-Yaus remains challenging, for this case we provide arguments based on our results for rigid Calabi-Yaus. }, implying $d g_s \simeq d N$. On the other hand, the distribution of the rank of the condensing gauge group in the string landscape is still poorly understood.\footnote{We are thankful to R. Savelli, R. Valandro and A. Westphal for illuminating discussions on this point.} Ref. \cite{Louis:2012nb} estimated the largest value of $\mathfrak{n}$ as a function of the total number of K\"ahler moduli, counted by the topological number $h^{1,1}$, but did not study how the number of vacua varies in terms of $\mathfrak{n}$. Moreover the F-theory analysis of \cite{Louis:2012nb} is based on the assumption that the formation of gaugino condensation in the low-energy 4D theory is not prevented by the appearance of unwanted matter fields.

In fact, as shown in \cite{Grassi:2014zxa, Morrison:2014lca}, F-theory sets severe constraints on the form of `non-Higgsable' gauge groups which guarantee that the low-energy theory features a pure super Yang-Mills theory undergoing gaugino condensation. Even if simple gauge groups like $SU(2)$ or $SU(3)$ are allowed, they do not survive in the weak coupling type IIB limit since they arise only from non-trivial $(p,q)$ 7-branes that do not admit a perturbative description in terms of D7-branes. The only type IIB case allowed for pure super Yang-Mills is $SO(8)$ which corresponds to $\mathfrak{n}=6$. This fits with the fact that all explicit type IIB Calabi-Yau orientifold models which have been constructed so far, feature exactly an $SO(8)$ condensing gauge group \cite{Cicoli:2011qg, Cicoli:2012vw, Cicoli:2013mpa, Cicoli:2013cha, Cicoli:2017shd}.

A non-perturbative superpotential can however arise also in a hidden gauge group with matter fields, even if there are constraints on the numbers of flavours and colours \cite{Affleck:1983rr}. Chiral matter can always be avoided by turning off all gauge fluxes on D7-branes but vector-like states are ubiquitous features of type IIB models obtained as the $g_s\to 0$ limit of F-theory constructions. Given that the interplay between vector-like states and the generation of a non-perturbative superpotential has not been studied in the literature so far, it is not clear yet if $\mathfrak{n}$ can only take two values, i.e. $\mathfrak{n}=1$ for ED3s and $\mathfrak{n}=6$ for a pure $SO(8)$ theory, or an actual $\mathfrak{n}$-distribution is indeed present in the string landscape. Even if we do not have a definite answer to this question at the moment, we can however argue that, if an actual $\mathfrak{n}$-distribution exists, the number of states $N$ is expected to decrease when $\mathfrak{n}$ increases since D7-tadpole cancellation is easier to satisfy for smaller values of $\mathfrak{n}$. We shall therefore take a phenomenological approach and assume $d N \sim -\mathfrak{n}^{-r} \,d \mathfrak{n}$ with $r>0$. 
Therefore (\ref{m32distr}) reduces to:
\be
d m_{3/2} \simeq \mathfrak{n}\,m_{3/2} \left[\ln\left(\frac{M_p}{m_{3/2}}\right)\right]^2  \left[1 - \frac{c_2 \,\mathfrak{n}^{r-2}}{\ln\left(\frac{M_p}{m_{3/2}}\right)} \right] d N\,.
\label{m32distrib}
\ee
For $0< r\leq 2$, the distribution of $m_{3/2}$ is therefore driven mainly by the distribution of the string coupling:
\be
\frac{d N}{d m_{3/2}} \simeq \frac{1}{\mathfrak{n}\,m_{3/2}} \left[\ln\left(\frac{M_p}{m_{3/2}}\right)\right]^{-2}\qquad\Rightarrow \qquad N_\LVS (m_{3/2}) \sim \ln\left(\frac{m_{3/2}}{M_p}\right)\,,
\label{m32Distr}
\ee
where we neglected subleading logarithmic corrections.\footnote{Notice that the result is unchanged if the distribution of the dilaton is taken to be power-law.} Comparing this results with (\ref{GravDistr}), we realise that in LVS models $\beta=-2$, and so we end up with the following the distribution of the gravitino mass:
\be 
\boxed{\,\rho_\LVS(m_{3/2}) \, \sim \, \frac{1}{\mathfrak{n}\,m_{3/2}^2} \left[\ln\left(\frac{M_p}{m_{3/2}}\right)\right]^{-2}}\,.
\ee
On the other hand, for $r>2$, the distribution of the number of D7-branes starts to play a r\^ole in the distribution of $m_{3/2}$ when $\mathfrak{n}$ is large. However, except for different subdominant logarithmic corrections, the leading order expression for the number of states as a function of the gravitino mass would still be given by (\ref{m32Distr}). It is reassuring to notice that our result is independent on the exact form of the unknown $\mathfrak{n}$-distribution.\footnote{This is true unless $N$ decreases exponentially when $\mathfrak{n}$ increases but this behaviour looks very unlikely. }

Notice that the result (\ref{m32distrib}) applies also to the distribution of the soft terms. In fact, as summarised in (\ref{SofttermsLVS}) and as reviewed more in detail in App. \ref{AppB}, the gravitino mass can generically be written in terms of the energy scale associated to the soft terms as $m_{3/2}\simeq M_{\rm soft}^{1/p}$ where for D7-branes $p=1$, while for D3-branes $p=2$ for gaugino masses and $p=2$ or $p=3/2$ for scalar masses depending on the `uplifting' mechanism. Thus in LVS models also the distribution of the soft masses turns out to be logarithmic:
\be
N_\LVS (M_{\rm soft}) \sim \frac{1}{p} \ln\left(\frac{M_{\rm soft}}{M_p}\right)\,.
\label{MsoftDistr}
\ee
This result is particularly important for models where the visible sector is realised on stacks of D3-branes since in this case the visible sector gauge coupling is set by $g_s$ which is therefore fixed by the phenomenological requirement of reproducing the observed visible sector gauge coupling. Hence the distribution of $m_{3/2}$ (or equivalently $M_{\rm soft}$) is entirely determined by the distribution of $\mathfrak{n}$. For this scenario, it would be very interesting to know if a non-perturbative superpotential can indeed be generated also in the presence of vector-like matter. If this does not turn out to be the case, then the value of the gravitino mass in LVS models with the visible sector on D3-branes can only take two values (setting the string coupling of order the GUT coupling $g_s = \alpha_{\scriptscriptstyle GUT} = 1/25$, $A_s\sim\mathcal{O}(1-10)$ and $\xi=1$):
\bi
\item \textbf{ED3-instantons}: in this case $\mathfrak{n}=1$ and:  
\be
m_{3/2} = g_s\,c_1\,e^{- \frac{c_2}{g_s}} \sim \mathcal{O}(10^{-26}-10^{-27})\,{\rm GeV}\,.
\ee

\item \textbf{Pure $SO(8)$}: in this case $\mathfrak{n}=6$ and:
\be
m_{3/2} = g_s\,\frac{c_1}{6}\,e^{- \frac{c_2}{6\,g_s}}\sim \mathcal{O}(10^9-10^{10})\,{\rm GeV} \,.
\ee
\ei
Notice that the ED3-case would be viable only for models where supersymmetry is broken by brane construction, so that the soft terms are at the string scale which is however around the TeV-scale. The extremely low value of $m_{3/2}$ might be helpful to control corrections to the vacuum energy coming from loops of bulk states \cite{Cicoli:2011yy}. The pure $SO(8)$ case instead corresponds to a more standard situation where however TeV-scale soft terms could be achieved only via sequestering effects \cite{Blumenhagen:2009gk, Aparicio:2014wxa}.

\subsection{KKLT models}

Let us now study the distribution of the gravitino mass in KKLT models where the minimum and $m_{3/2}$ are given respectively by (\ref{KKLTmin}) and (\ref{m32KKLT}). Varying the gravitino mass with respect to the two flux-dependent parameters $g_s$ and $|W_0|$, and the integer parameter $\mathfrak{n}$, we obtain:
\be
d m_{3/2} \simeq m_{3/2}\left(\frac{d |W_0|}{|W_0|}+\frac12\frac{d g_s}{g_s}-\frac{3}{2}\frac{d \mathfrak{n}}{\mathfrak{n}} \right),
\label{dm32KKLT}
\ee
where we neglected the subleading variation of the logarithm. Following the arguments given in Sec. \ref{LVSsection} and in App. \ref{gsdistribution}, we assume a uniform distribution of the string coupling, i.e. $d N \simeq d g_s $, and a phenomenological scaling of the distribution of $\mathfrak{n}$ of the form $d N\simeq -\mathfrak{n}^{-r}\,d \mathfrak{n}$. Moreover the distribution of $W_0$ as a complex variable is also uniform \cite{Denef:2004ze}, resulting in $d N \simeq |W_0| d |W_0|$. Thus (\ref{dm32KKLT}) reduces to:
\bea
d m_{3/2} &\simeq & m_{3/2}\left(\frac{1}{|W_0|^2}+\frac{1}{2g_s}+\frac32\,\mathfrak{n}^{r-1} \right) d N \nn \\
&\simeq& \frac{M_p^2}{m_{3/2}} \left[\frac{g_s}{\mathfrak{n}^3 |\ln W_0|^3} +\frac{\epsilon^2}{2}\left(\frac{1}{g_s}+ 3 \mathfrak{n}^{r-1}\right) \right] d N\,,
\label{GravDistrKKLT}
\eea
where $\epsilon \equiv m_{3/2}/M_p$. In order to trust the effective field theory description we need to require $\epsilon \ll 1$, which implies that the distribution of the gravitino mass is dominated by the first term in (\ref{GravDistrKKLT}), i.e. by the distribution of the flux superpotential:
\be
\frac{d N}{d m_{3/2}} \simeq \left(\frac{\mathfrak{n}^3 |\ln W_0|^3}{g_s}\right) \frac{m_{3/2}}{M_p^2} \simeq \frac{m_{3/2}}{M_p^2} \qquad \Rightarrow \qquad N_\KKLT(m_{3/2}) \sim \left(\frac{m_{3/2}}{M_p} \right)^2.
\label{NKKLT}
\ee
Comparing this results with (\ref{GravDistr}), we realise that in KKLT models $\beta=0$, in agreement with previous predictions \cite{Douglas:2004qg}. Thus we end up with the following the distribution of the gravitino mass:
\be 
\boxed{\,\rho_\KKLT(m_{3/2}) \, \sim \, \frac{1}{M_p^2} \left(\frac{\mathfrak{n}^3 |\ln W_0|^3}{g_s}\right)\,\sim \,\text{const.}}
\label{RhoKKLT}
\ee
As reviewed App. \ref{AppB}, in KKLT models the soft terms are proportional to the gravitino mass (up to a possible 1-loop suppression factor for visible sector models on D3-branes). Therefore (\ref{NKKLT}) and (\ref{RhoKKLT}) give also the distribution of the soft terms in KKLT models.

\subsection{Perturbatively stabilised models}

Let us now study the distribution of the gravitino mass in perturbatively stabilised models where the minimum and $m_{3/2}$ are given respectively by (\ref{PertMin}) and (\ref{m32Pert}). Varying the gravitino mass with respect to the three flux-dependent parameters $g_s$, $|W_0|$ and $c$, we obtain:
\be
d m_{3/2} \simeq  m_{3/2}\left(\frac{d |W_0|}{|W_0|}-4\,\frac{d g_s}{g_s} -3\,\frac{d c}{c}\right),
\label{dm32Pert}
\ee
As discussed in \cite{Denef:2004ze} and in App. \ref{gsdistribution}, both $g_s$ and $W_0$ are expected to be uniformly distributed, and so we take $d N \simeq d g_s$ and $d N \simeq |W_0| d |W_0|$. Moreover, as stressed in Sec. \ref{PertSec}, the coefficient $c$ is a function of the complex structure moduli which are fixed in terms of flux quanta, and so it is naturally expected to be of order $c\sim \mathcal{O}(1-10)$. However the minimum in (\ref{PertMin}) lies at $\vo\gg 1$ only if the flux quanta are tuned such that $c \sim g_s^{-(3/2+q)}\gg 1$ with $q>0$. Given that this is a tuned situation, we expect the number of vacua at $c \gg 1$ to be suppressed with respect to the region with $c\sim \mathcal{O}(1-10)$. This behaviour is well described by a distribution of $c$ with a phenomenological scaling of the form $d N\simeq - c^{-k}\,d c$ with $k>0$. Using all these relations, (\ref{dm32Pert}) becomes:
\bea
d m_{3/2} &\simeq&  m_{3/2}\left(\frac{1}{|W_0|^2}-\frac{4}{g_s} + 3\,c^{k-1}\right) d N \nn \\
&\simeq& m_{3/2}\left(3\,c^{k-1}- \frac{4}{g_s}\right) d N\,,
\label{GravDistrPert}
\eea
where we focused on the region with $|W_0|\sim \mathcal{O}(1-10)$ and $g_s\lesssim 0.1$. Notice that for such a small value of the string coupling and $0<k\leq 1$, the second term in (\ref{GravDistrPert}) would dominate over the first one. However this is a regime where the distribution of the coefficient $c$ would be almost uniform, and so $c$ would be in the regime $c\sim \mathcal{O}(1-10)$ where the effective field theory is not under control. We focus therefore on $k>1$ where the distribution of $c$ starts to deviate from begin uniform, signaling that $c$ is tuned to large values. In this case the distribution of the gravitino mass is dominated by the first term in (\ref{GravDistrPert}) and becomes:
\be
\frac{d N}{d m_{3/2}} \simeq \frac{1}{m_{3/2}\,\,c^{k-1}}\simeq \left(\frac{g_s^4}{|W_0|}\right)^{\frac{(k-1)}{3}} \,\frac{1}{M_p}\,\left(\frac{m_{3/2}}{M_p}\right)^{\frac{(k-4)}{3}}\,,
\ee
which implies:
\be
N_\PERT(m_{3/2}) \sim \left(\frac{m_{3/2}}{M_p}\right)^{\frac{(k-1)}{3}}\,.
\label{NPert}
\ee
Comparing this results with (\ref{GravDistr}), we realise that in perturbatively stabilised models $\beta=(k-7)/3$. Hence we end up with the following distribution of the gravitino mass:
\be 
\boxed{\,\rho_\PERT(m_{3/2}) \, \sim \, \frac{1}{M_p^2}\,\left(\frac{m_{3/2}}{M_p}\right)^{\frac{(k-7)}{3}}}\,.
\label{RhoPert}
\ee
This result is qualitatively similar to the one of KKLT models (which are reproduced exactly for $k=7$), showing that scenarios where the K\"ahler moduli are stabilised by perturbative effects favour higher values of the gravitino mass. This behaviour is somewhat expected since these models, similarly to KKLT, can yield trustable vacua only relying on tuning the underlying parameters. This tuning, in turn, reflects itself on the preference for larger values of $m_{3/2}$. As mentioned in Sec. \ref{PertSec}, in perturbatively stabilised models the soft terms are expected to be proportional to the gravitino mass, and so (\ref{NPert}) and (\ref{RhoPert}) give also the distribution of the soft terms in these models.

\section{Discussion}
\label{Sec4}

In this section we summarise our results and discuss them in the context of the original results of \cite{Ashok:2003gk, Douglas:2004qg, Douglas:2004zg, Denef:2004dm, Denef:2004ze, Denef:2004cf}, as well as the subsequent results obtained in \cite{Dine:2004is, Dine:2005iw, Dine:2005yq, Dine:2004ct}.

\subsection{Interplay with previous results}

Firstly, we have stressed in Sec. \ref{Sec2} that K\"ahler moduli stabilisation is a critical requirement for a proper treatment of the statistics of supersymmetry breaking. The reason is that a stable solution requires the F-terms of the axio-dilaton and the complex structure moduli to be suppressed with respect to the F-terms of the K\"ahler moduli. The statistics of supersymmetry breaking is thus entirely driven by the F-terms of the K\"ahler moduli at their stabilised values. 

As we have shown, the no-scale structure at tree level has important consequences for the statistics of supersymmetry breaking. It implies that in order to
obtain vacua where the $\alpha'$ and $g_s$ expansions are under control, terms in the effective action which are part of separate expansions have to be balanced against each other (see \cite{Cicoli:2018kdo} for a detailed discussion of this point). For example, in LVS we find that $\alpha'$ corrections associated with the overall volume are balanced against a non-perturbative correction associated with a blow-up modulus. In KKLT, on the other hand, non-perturbative effects are balanced against an exponentially small flux superpotential. This implies that the stabilisation mechanism pushes us to particular regions in moduli space -- in LVS the overall volume is large, while in KKLT $|W_0|$ is inevitably small -- where the gravitino mass takes specific values.

This has important implications for the statistics of soft terms which in gravity mediation are determined by $m_{3/2}$. As we have seen in Sec. \ref{Sec3}, different stabilisation mechanisms predict different distributions of the gravitino mass (and hence the soft terms) in the landscape. This is due to the fact that different no-scale breaking effects used to fix the K\"ahler moduli lead to a different dependence of $m_{3/2}$ on the flux-dependent microscopic parameters $W_0$, $g_s$ and $c$ whose distribution (together with the one of $\mathfrak{n}$) ultimately governs the statistics of the soft terms, as is evident from (\ref{m32distr}), (\ref{dm32KKLT}) and (\ref{GravDistrPert}). In particular, we found that in LVS models the distributions of the gravitino mass and soft terms are logarithmic, as shown in (\ref{m32Distr}) and (\ref{MsoftDistr}). On the other hand, for KKLT and perturbative stabilisation, the distributions are power-law, as shown in (\ref{NKKLT}) and (\ref{NPert}). The difference in behaviour comes from the fact that in the LVS case one has from (\ref{m32LVS}):
\be
m_{3/2} \sim M_p\,e^{- \frac{1}{g_s}}\,,
\label{m32simple}
\ee
which, when combined with the fact that $g_s$ is uniformly distributed as shown in App. \ref{gsdistribution}, yields a logarithmic distribution for $m_{3/2}$. For KKLT, one has instead from (\ref{m32KKLT}):
\be
m_{3/2} \sim  |W_0|\, M_p\,,
\ee
which results in a power-law distribution of the gravitino mass since since $|W_0|$ is uniformly distributed. A similar reasoning applies in the case of perturbative stabilisation.

Interestingly, we note that both power-law \cite{Douglas:2004qg, Denef:2004ze, Denef:2004cf} as well as logarithmic distributions \cite{Dine:2004is, Dine:2005iw, Dine:2005yq, Dine:2004ct} have been obtained by different groups in the literature, albeit for reasons different from the ones we have derived. The power-law distribution of gravitino masses in (\ref{NKKLT}) and (\ref{NPert}) for KKLT and perturbatively stabilised vacua reproduces the results of \cite{Douglas:2004qg, Denef:2004ze, Denef:2004cf} which were based on the assumption of a democratic distribution of complex structure F-terms caused by the uniform distribution of $|W_0|$, as we have reviewed in Sec. \ref{SUSYstatRev}. In KKLT and perturbatively stabilised vacua, the supersymmetry breaking scale is instead determined by the F-terms of the K\"ahler moduli but we obtain the same behaviour given that in these two K\"ahler stabilisation schemes they are also governed dominantly by $|W_0|$. On the other hand, the logarithmic distributions (\ref{m32Distr}) and (\ref{MsoftDistr}) of LVS models reproduce the results of \cite{Dine:2004is, Dine:2005iw, Dine:2005yq, Dine:2004ct} whose derivation was based on the general nature of dynamical supersymmetry breaking: if the scale of supersymmetry breaking is given by $m_{3/2} \sim  M_p\,e^{-8\pi^2/g^2}$ with a flat distribution in the coupling $g^2$, then $m_{3/2}$ would obey a logarithmic distribution. Indeed, this expectation is exactly reproduced by the expression (\ref{m32simple}) for the gravitino mass in LVS models since in type IIB compactifications the gauge coupling $g$ of a hidden sector supporting non-perturbative effects which break supersymmetry dynamically scales as $g^2 \sim g_s$.    

Determining which distribution, power-law or logarithmic, is more representative of the structure of the flux landscape therefore translates into the question of which vacua with stabilised K\"ahler moduli arise more frequently. Given that LVS models can be realised for natural values of the vacuum expectation value of the flux superpotential, $|W_0|\sim \mathcal{O}(1-10)$, while KKLT models can be constructed only via tuning $|W_0|$ to exponentially small values (similar considerations about tuning of the underlying parameters apply also to perturbatively stabilised vacua), we tend to conclude that the distribution of the scale of supersymmetry breaking seems to be logarithmic. However, more detailed studies are needed in order to find a precise definite answer to this important question (see \cite{DeWolfe:2004ns, Cicoli:2013cha, Demirtas:2019sip, Cole:2019enn} for initial studies on the determination of the number of vacua as a function of $|W_0|$ and $g_s$).

Finally, we would like to make a few comments discussing our results in the context of the cosmological constant. The explicit analysis carried out in the previous section focused on solutions with zero cosmological constant and so far we considered the joint distribution of the supersymmetry breaking scale and the cosmological constant. As we have mentioned before, soft masses for the SM sector are typically predominantly determined by a small set of non-vanishing F-terms and D-terms in the theory. On the other hand, the cosmological constant receives contributions from {{\it all}} F and D-terms, many of which can be sequestered from the SM sector and make subdominant contributions to supersymmetry breaking. This has two implications: $(i)$ to compute distributions of the cosmological constant one needs to have a knowledge of all the uplift contributions, which is generally challenging; and $(ii)$ since a large number of contributions to the cosmological constant do not affect the soft masses, one can expect the distribution of the cosmological constant to be independent of the distribution of the soft masses. LVS models are a neat example where the decoupling between the statistics of supersymmetry breaking and the cosmological constant emerges clearly. In fact, combining the expression (\ref{VLVS}) of the scalar potential of LVS models with the location of the minimum (\ref{LVSmin}), it is easy to see that the depth of the non-supersymmetric AdS vacuum is:
\be
V_\LVS \sim - \frac{|W_0^2|}{\vo^2}\sim - m_{3/2}^3\,M_p\,.
\ee
This implies that any hidden sector whose dynamics is responsible for dS uplifting has to provide a contribution to the scalar potential whose order of magnitude is:
\be
V_{\rm up} \sim |F_{\rm hid}|^2 \sim m_{3/2}^3\,M_p\,.
\ee
In turn this hidden sector generates a contribution to the soft terms via gravity mediation which is suppressed with respect to the gravitino mass:
\be
\delta M_{\rm soft} \sim \frac{F_{\rm hid}}{M_p} = m_{3/2}\sqrt{\frac{m_{3/2}}{M_p}} \ll m_{3/2}\,.
\ee
Hence, if the F-terms of other hidden sectors (like for example the F-term of the K\"ahler modulus controlling the volume of the 4-cycle wrapped by the SM stack of D7-branes) generate soft terms of order $m_{3/2}$, the contribution from $F_{\rm hid}$ is clearly negligible. Notice that this implies that the distribution of the supersymmetry breaking scale is the same at least for all vacua with cosmological constant in the range $\pm V_\LVS$. Of course, the distribution could change if we consider vacua with much higher values of the cosmological constant.
 
\subsection{Implications for phenomenology}

We now turn to a brief discussion of the implications of our findings for low energy phenomenology. The ATLAS collaboration has provided 95\% CL search limits for gluino pair production within various simplified models using data sets that vary from 36-139 fb$^{-1}$ at $\sqrt{s}=13$ TeV \cite{Aaboud:2017vwy}. The approximate bound from these searches is that $m_{\tilde{g}}\gtrsim 2.2$ TeV. The limits coming from CMS are comparable \cite{Vami:2019slp}. Searches for top squark pair production yield the limit $m_{\tilde{t}}\gtrsim 1$ TeV \cite{ATLAS:2019oho, Sirunyan:2019glc}. 

We have found that the statistics of type IIB flux vacua generally prefers a draw towards high scale supersymmetry: a mild logarithmic draw in the case of LVS, and a strong power-law draw in the case of KKLT and perturbatively stabilised vacua. Given the current limits on gluinos and squarks, can one surmise that it is this statistical draw that is being played out at experiments?

Of course, the problem with this interpretation is that high scale supersymmetry breaking leads to fine-tuning issues for the mass of the Higgs, obviating, at least from the low-energy perspective, the introduction of supersymmetry as a solution to the gauge hierarchy problem in the first place. The severity of this issue may be quantified by the choice of suitable fine-tuning measures. In other words, since stringy naturalness (the bias towards a property favored by vacuum statistics, in this case, high scale supersymmetry breaking) leads one to posit heavier superpartners, this tendency should somehow be mitigated by a fine-tuning penalty as one goes to higher scales. But  which fine-tuning measure should one use, and how much penalty should one impose? 

The widely adopted Barbieri-Giudice measure \cite{Barbieri:1987fn} is defined as $\Delta_\BG\equiv {\rm max}_i |\frac{\partial \ln m_Z^2}{\partial\ln p_i}|$ with, for example, $\Delta_\BG<10$ corresponding to $\Delta_\BG^{-1}=10\%$ fine-tuning. The $p_i$ are the fundamental parameters of the theory, while $m_Z$ denotes the mass of the $Z$ boson. Taking the parameters to be the various soft terms and $\mu$ parameter from the mSUGRA/CMSSM model and requiring $10\%$ 
fine-tuning, one obtains upper limits of $m_{\tilde{g}}\sim 400$ GeV \cite{Baer:2020kwz}. Most other superpartners  are also close to the weak scale (defined as $m_{\rm weak}\simeq m_{W,Z,h}\sim 100$ GeV). It is thus clear from the Barbieri-Giudice measure that supersymmetry is already very finely tuned from LHC data. From the perspective of the landscape, one can impose a penalty on $\Delta_\BG$ for vacua with very high scale supersymmetry breaking (while also allowing for the fine-tuning indicated by data) but it is not entirely clear what the penalty should be or how to motivate it.

An alternative approach is to use anthropic arguments to motivate fine-tuning penalties on vacua with high scale supersymmetry breaking \cite{Baer:2017uvn, Baer:2019xww, Baer:2015rja}.\footnote{Indeed, the landscape is already a fertile arena where such arguments have been used in the past, most famously in the context of the cosmological constant problem \cite{Bousso:2000xa, Weinberg:1987dv}.} The atomic principle \cite{Agrawal:1997gf} comes closest in relevance in this context. It can be incorporated within the fine-tuning measure introduced in \cite{Baer:2012up}, whose starting point is the expression for the mass of the $Z$ boson in supersymmetry: $m_Z^2/2 \simeq -m_{H_u}^2-\mu^2-\Sigma_u^u(\tilde{t}_{1,2})$ (for details and exact expressions, we refer to the original paper and \cite{Baer:2020kwz}). Here, $\Sigma_u^u$ contains the various radiative corrections \cite{Baer:2012cf}. The fine-tuning penalty in this case posits that no single contribution in the expression for $m_Z$ can be too much larger than any other. This is quantified by the measure $\Delta_\EW$ which is the maximum among the quantities on the right hand side divided by the $m_Z^2/2$.

It is now clear how the atomic principle naturally plays into the fine-tuning measure $\Delta_\EW$. Given that the mass of the $Z$ boson is bounded by the atomic principle, one obtains an anthropic bound on the scale of the superpartners stemming from their contributions to the radiative corrections encapsulated in $\Sigma_u^u$. Indeed, requiring that the mass of the $Z$ boson should not exceed its measured value by a factor of 4 imposes $\Delta_\EW \lesssim 30$, which in turn translates into upper bounds on superpartner masses entering through the radiative corrections $\Sigma_u^u$. 

One thus has a logarithmic or power-law distribution of vacua biasing towards high supersymmetry breaking scales, tempered by a penalty of $\Delta_\EW \lesssim 30$ coming from the atomic principle. For power-law distributions, this leads to several predictions for superpartner masses that may be probed at the HL-LHC. For example, the statistical distribution for gluinos and top squarks are peaked around 4 TeV and 1.5 TeV, respectively. Suggestively, the Higgs mass appears to be peaked around 125 GeV for power-law distributions. A logarithmic distribution from the landscape, on the other hand, would imply that a low scale of supersymmetry breaking is reasonably probable, perhaps without relying too strongly on  anthropic arguments. The value of the weak scale may simply be a mild accident in that case. We leave a more detailed treatment of the phenomenology of the logarithmic case  for future work.

\section{Conclusions}
\label{Conclusions}

Understanding the distribution of the supersymmetry breaking scale in string vacua is an important question which can potentially have deep phenomenological implications. In this paper, we have revisited this question in the context of IIB flux vacua. In the first part of the paper, we argued that the details of  K\"ahler moduli stabilisation are absolutely necessary to study the distribution of the supersymmetry breaking scale. We then went on to study the distribution of the supersymmetry breaking scale (primarily focusing on vacua with zero cosmological constant) in three scenarios for K\"ahler moduli stabilisation: ($i$) models with purely non-perturbative stabilisation like in KKLT vacua; ($ii$) models where the K\"ahler moduli are frozen by balancing perturbative against non-perturbative effects as in LVS models; and ($iii$) models with purely perturbative stabilisation. For KKLT and models with perturbative stabilisation we found a power law distribution, while for LVS we found a logarithmic distribution. The logarithmic distribution is particularly interesting as it could well mean that we should remain optimistic about discovering superpartners in collider experiments.

Let us mention that our results for the distribution of the supersymmetry breaking scale in the type IIB flux landscape are based on the fact that $|W_0|$ and $g_s$ are uniformly distributed.\footnote{The result for LVS is unchanged as long as the distribution for $g_s$ is a power-law.} While in the literature there is a lot of evidence in favour of this assumption (as we also have shown for the distribution of the string coupling for rigid Calabi-Yaus), more detailed numerical studies are needed in order to confirm the validity of this behaviour for the general case. This investigation is crucial also to determine which distribution, power-law or logarithmic, is predominant in the flux landscape since the distribution of the vacuum expectation value of the flux-generated superpotential is a key input for determining the relative preponderance of KKLT and LVS vacua. 

This work opens up several interesting directions for future research. Firstly, it is important to carry out a detailed study along the lines of \cite{Baer:2020kwz} to understand the phenomenological implications of the logarithmic distribution. In order to make contact with observations it will be crucial to incorporate also bounds arising from the cosmological context (such as the cosmological moduli problem). Our analysis has focused on a small (but highly attractive from the point of view of phenomenology) corner of the string landscape, i.e type IIB flux compactifications on Calabi-Yau orientifolds. It will be interesting to carry out an analysis in the same spirit as this paper in other corners of the landscape.\footnote{Even within the context of type IIB, it will be interesting to explore the constructions in \cite{Saltman:2004jh} which naturally have a high scale of supersymmetry breaking, even if the visible sector phenomenology is not well developed in this setting.} A related but very challenging question is to investigate if early universe cosmology gives us a natural measure on the space of solutions in string theory.

\section*{Acknowledgements}

We would like to thank S. Ashok, H. Baer, R. Savelli, G. Shiu, R. Thangadhurai, R. Valandro and A. Westphal for useful conversations.

\appendix 

\section{Distribution of the string coupling}
\label{gsdistribution}

In this appendix we discuss the distribution of $g_s$ in type IIB flux compactifications. This has been studied in \cite{Ashok:2003gk, Denef:2004ze},
and we follow here their analysis to obtain an understanding of the distribution in the region of our interest, i.e. low values of $g_s$. As in \cite{Ashok:2003gk, Denef:2004ze}, we will carry out a detailed numerical analysis for the simple tractable case of rigid Calabi-Yaus, and use these results to develop intuition for general Calabi-Yaus.

For rigid Calabi-Yaus, the $\tau$ modulus ($\tau = a + {i \over g_s}$, where $g_s$ is the dilaton and $a$ its axionic partner), has a linear superpotential:
\bel{wk}
   W = A \tau + B\,,
\ee
the `fluxes'  $A = a_1 + i a_2$ and $B = b_1 + i b_2$ take values in ${\mathbb{Z} + i \mathbb{Z}}$. The tadpole cancellation condition is:
\bel{tad}
   \text{Im} (A^{*} B) = L \equiv {\rm{Det}}(X) =L\,,
\ee
where $X$ is the matrix:
\bel{Xmat}
  X =
\begin{pmatrix}
  a_1 &a_2 \\
  b_1  & b_2
\end{pmatrix}.
\ee
The form of the tadpole condition in \pref{tad} makes it manifest that the tadpole cancellation condition has an $SL(2, {\mathbb{Z}})$ symmetry, i.e. transformations of the form:
\bel{tadtran}
  X \to X' = M X\,,
\ee
map solutions to solutions, with M $\in$ $\slz$. Taking the matrix $M$ to be: 
$$
M= 
\begin{pmatrix}
p & q \\
r & s
\end{pmatrix},
$$
the explicit form of the the transformation is given by:
\bel{exptra}
  \begin{pmatrix}
  a'_1 &a'_2 \\
  b'_1  & b'_2
\end{pmatrix}
= 
\begin{pmatrix}
p & q \\
r & s
\end{pmatrix}
.
\begin{pmatrix}
  a_1 &a_2 \\
  b_1  & b_2
\end{pmatrix}
= 
\begin{pmatrix}
pa_1 + qb_1 & pa_2 + qb_2 \\
ra_1 + sb_1 &  ra_2 + sb_2 
\end{pmatrix}.
\ee
Now, let us come to the vacua. They are supersymmetric:
\bel{vac}
DW = 0  \leftrightarrow  \bar{\tau} = - {B \over A} \implies  \tau = {{- b_1 + ib_2} \over {a_1 - ia_2}}\,.
\ee
Note that under the above described $\slz$ transformation:
\bea
\label{ztran}
\tau \to \tau' = {{- b'_1 + ib'_2} \over {a'_1 - ia'_2}} ={{s \tau - r }\over  {-q \tau +p}}\,.
\eea
This is an $\slz$ action on $\tau$ associated with the matrix $Y$ given by:\footnote{The fact that $Y$ is an element of $\slz$ follows from the fact that its determinant is the same as the one of $M$.}
\bel{tmatr}
Y=
\begin{pmatrix}
s & -r \\
-q & p
\end{pmatrix}
\ee
Therefore, given the $\slz$ symmetry of type IIB, the action does not generate physically distinct solutions.\footnote{There is another $\slz$ symmetry of the equation \pref{tad}. This involves taking $X \to X.N $, where $N$ is an $\slz$ matrix. It is easy to see that such transformations do not correspond to $\slz$ transformations of $\tau$. In this case, $a_i \to a_{l} N_{lk}$ and $b_j \to b_l N_{lj}$. Thus we can start with a point with $a_1,b_1 \neq 0$ and $a_2, b_2 = 0$ (i.e. $\tau$ on the real axis) and map it to a point where $a_1,a_2, b_1, b_2 \neq 0$. Thus a point on the real line can get mapped to a point in the interior of the upper half plane. Thus, this does not correspond to an $\slz$ transformation of $\tau$. Hence, this cannot be thought of as a `gauge' transformation.} In fact, solutions related by this symmetry should be considered as equivalent.

The above described gauge symmetry is crucial to understand the solution space. Firstly, we can use the $\slz$ symmetry to set $a_2 =0$. This implies: 
\bel{bz}
 {\tau} = - { b_1 \over a_1} + i { b_2 \over a_1}\,.
\ee
Also, the tadpole condition reduces to:
\bel{newtad}
a_1 b_2 = L\,.
\ee
Requiring $\rm{Im} (\tau) >0$, yields:
\bel{solc}
{b_2 \over a_1} > 0 \implies {b^2_2 \over a_1 b_2 }  \implies  {b^2_2 \over L  }  \implies L  > 0\,.
\ee
Thus, we have the condition $L  = a_1 b_2$ with $ L > 0$. Hence, $a_1$ and $b_2$ have to be integers which divide $L$ with $L > 0$. To see what values $b_1$ can take, we need to examine the residual $\slz$ invariance. The residual $\slz$ transformations correspond to transformations which maintain the condition $a_2 = 0$, from \pref{exptra} we see that this implies that $q=0$. Thus the $\slz$ matrix must take the form:
\bel{ressl}
      \begin{pmatrix}
 1 & 0 \\
 r & 1
 \end{pmatrix} 
 \ee
where $r$ is an integer.\footnote{Note that
  $
    \begin{pmatrix}
 1 & 0 \\
 r & 1
 \end{pmatrix}
 \equiv 
     \begin{pmatrix}
- 1 & 0\\
 -r & - 1
 \end{pmatrix}   $
 Hence we do not have to mod out by matrices of the form in the RHS of the equivalence.}
Now the action of an $\slz$ matrix of the form \pref{ressl} takes $b_1$ to:
\bel{resbtt}
    b_1 \to r a_1  + b_1\,.
\ee
This implies that $b_1$ takes the values $0, 1, ....|a_1-1|$. In summary, the analysis of \cite{Ashok:2003gk, Denef:2004ze} implies that vacua are characterised by:
\ben
\item An integer $a_1$ which divides $L$.  

\item For every such integer $b_1$ takes the values $0, 1, ....|a_1-1|$.

\item $b_2 =  {L \over a_1}$
 
\item The value of $\tau$ is given by:
\bel{bzf}
 {\tau} = - { b_1 \over a_1} + i { b_2 \over a_1}
\ee
\een

To get the distribution in the fundamental domain one takes the value of $\tau$ obtained from \pref{bzf} and maps it to the fundamental domain of $\slz$. This involves the repeated action of the generators:
\bel{slgenar}
   T: \tau \to \tau +1, \phantom{abcd}S:  \tau \to - {1 \over \tau}\,.
 \ee
The algorithm to bring a general point which is outside the fundamental domain to inside the fundamental domain is as follows: First, by repeated action of $T$ (or $T^{-1}$) the point is brought to the region $-\hf \leq Re(\tau) < \hf$. If this  process also brings the point to inside the fundamental  domain, then the algorithm terminates. Otherwise, one acts with the generator $S$. If this does not bring the point inside the fundamental domain one iterates the process of repeated action of $T$ (or $T^{-1}$) and a single action of $S$ (if needed) until the point is mapped to the fundamental domain. 

\begin{figure}[h]
\centering
 \begin{minipage}[b]{0.35\textwidth}
  \centering
  \includegraphics[width=\textwidth]{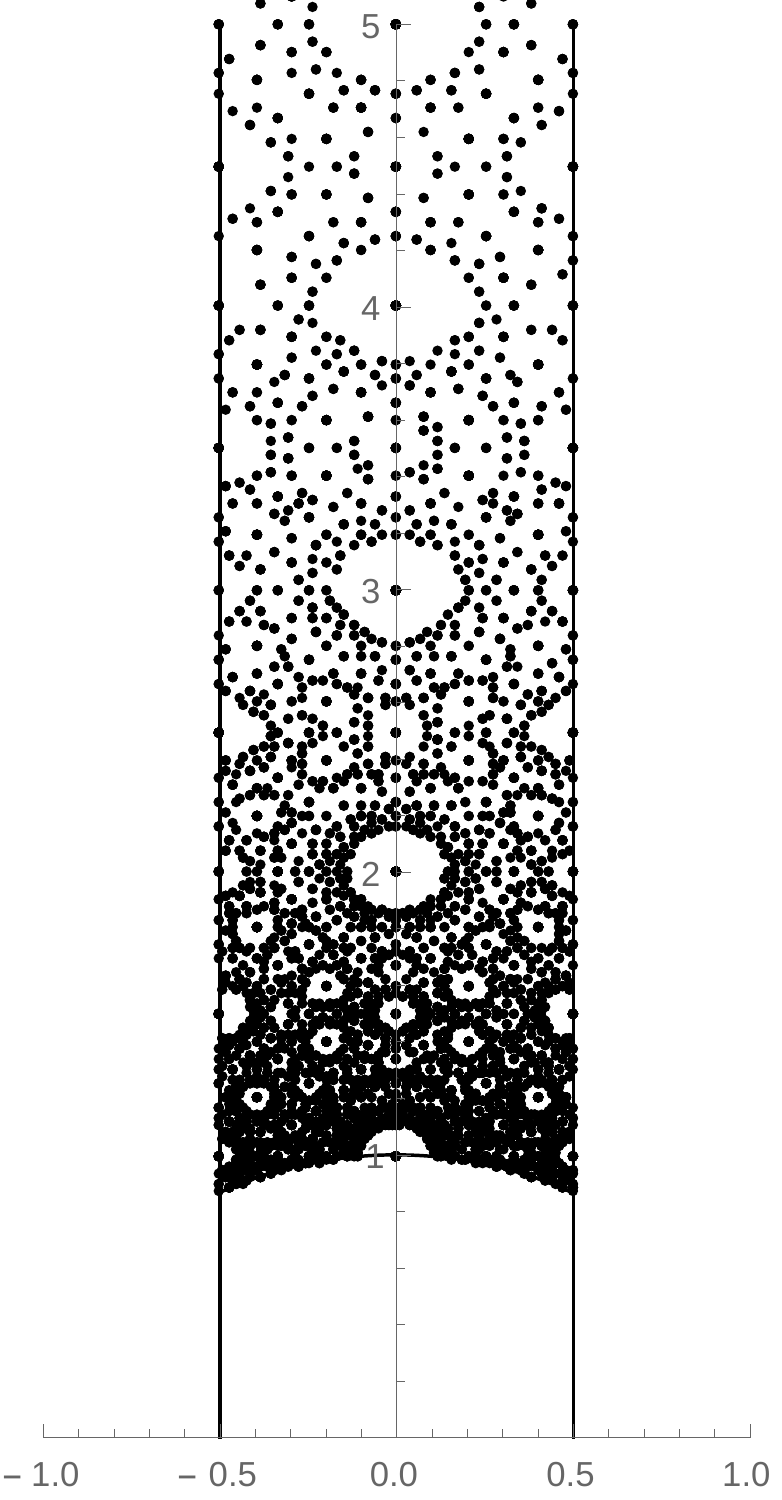}
  \caption{Values of $\tau$ for $L=100$.}
  \label{fig1}
 \end{minipage} 
 \hfill
 \begin{minipage}[b]{0.60\textwidth}
    \includegraphics[width=\textwidth]{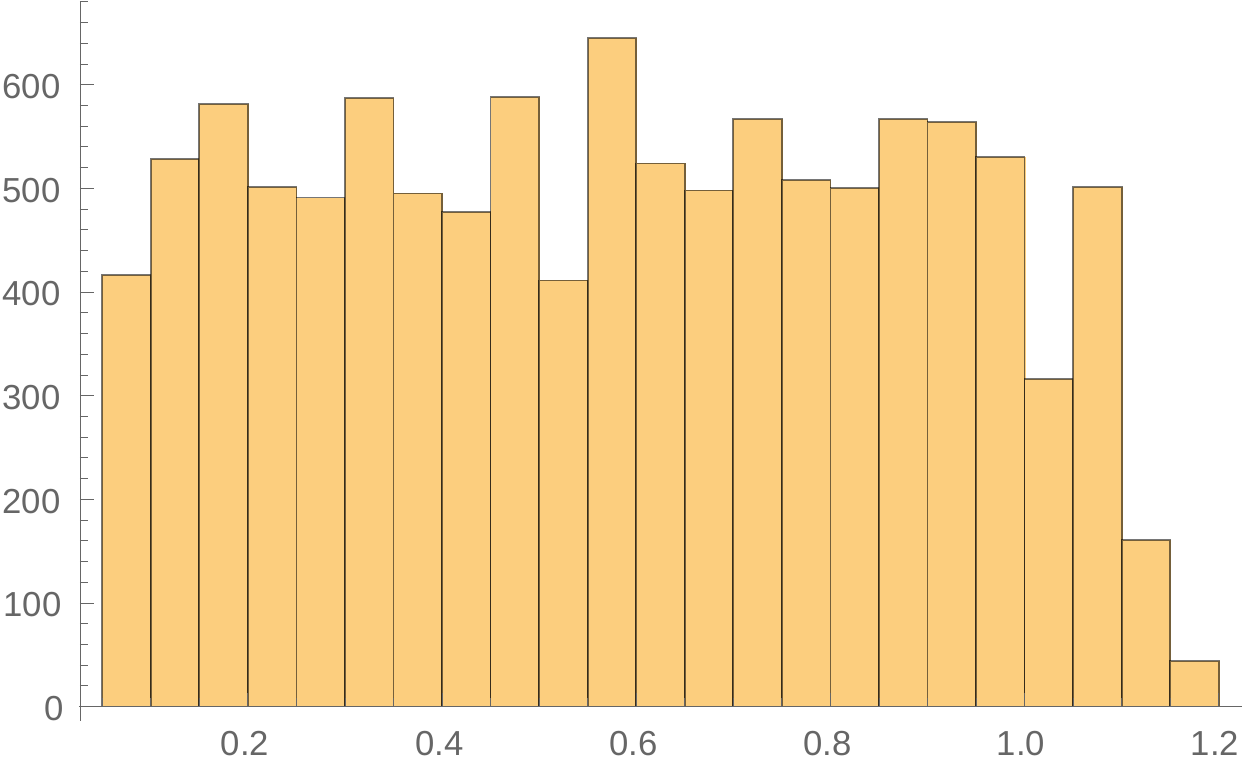}
    \caption{Distribution of $g_s$ for $L=100$.}
    \label{fig2}
 \end{minipage}
\end{figure}

Now, let us come to our discussion of the distribution of $g_s$. Note that the characterisation of inequivalent solutions implies that the values of imaginary part of $\tau$ as obtained in \pref{bzf} are bounded by:
\bel{bnd}
       {1 \over L} \leq \rm{Im} (\tau) \leq L\,.
\ee
It is easy to check that this condition is preserved by the algorithm to bring the points inside the fundamental domain. Thus the lowest value of $g_s$ is ${1 \over L}$. We have carried out detailed numerical studies to probe the distribution for small values of $g_s$ (in the region of phenomenological interest). First, we present  the results of our numerics for $L=100$. The distribution of $\tau$ in the fundamental domain is shown in Fig. \ref{fig1} and the distribution of $g_s$ is shown in Fig. \ref{fig2}. The results are consistent with that of \cite{Denef:2004ze}.

 \vspace{10pt}
 
The plot in Fig. \ref{fig2} shows that the distribution is roughly uniform for $g_s >  0.01$. 

\begin{figure}[h]
\centering
 \begin{minipage}[b]{0.45\textwidth}
  \centering
  \includegraphics[width=\textwidth]{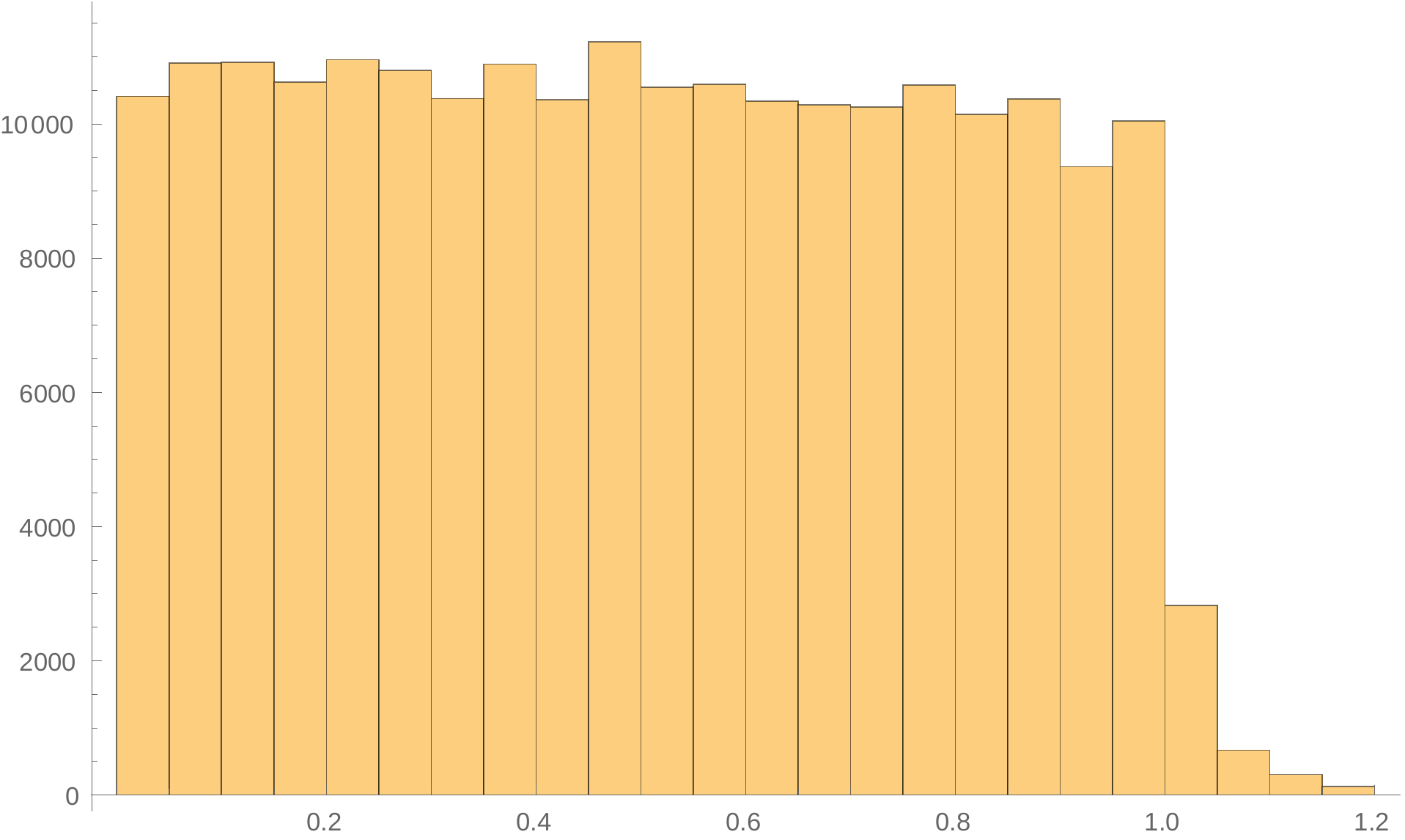}
  \caption{Distribution of $g_s$ for $L=500$.}
  \label{fig4}
 \end{minipage} 
 \hfill
 \begin{minipage}[b]{0.5\textwidth}
   \includegraphics[width=\textwidth]{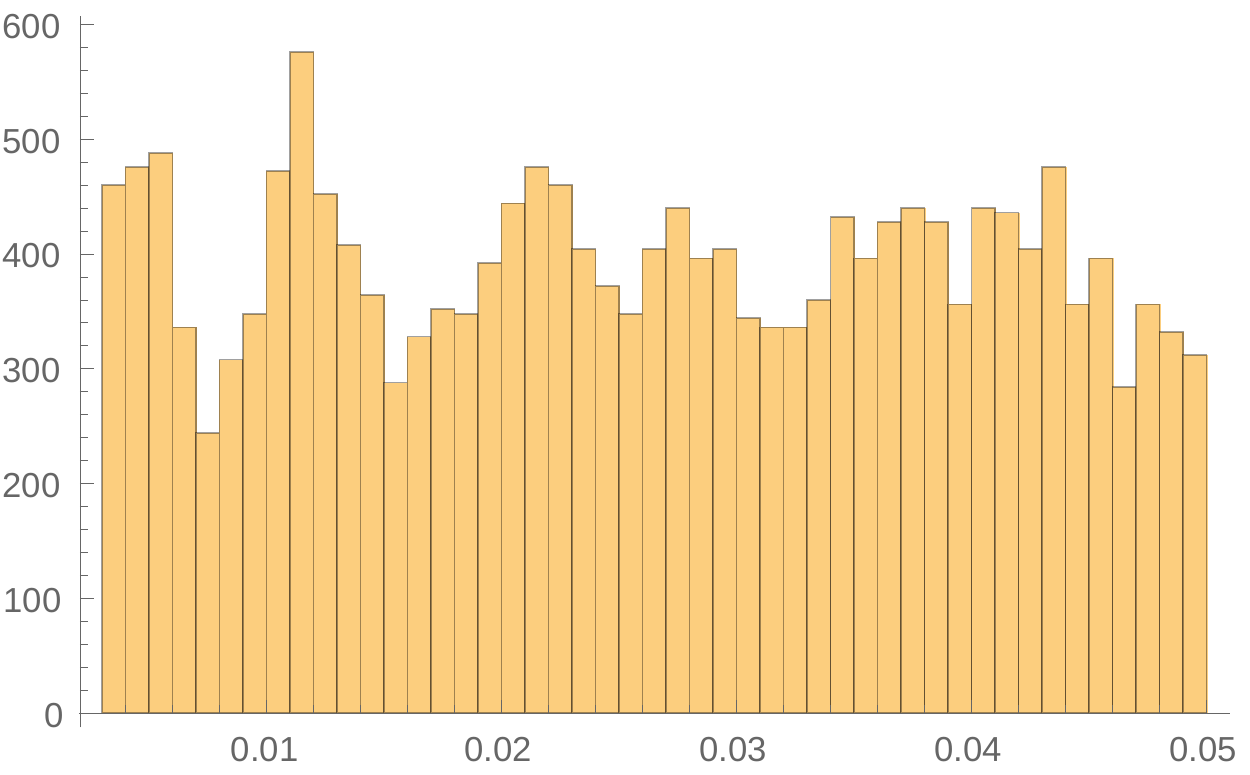}
 \caption{Distribution for small $g_s$ with $L$=500.}
 \label{fig3}
 \end{minipage}
\end{figure}

Next we present our results for $L= 500$. The distribution of the number of vacua as a function of $g_s$ is shown in Fig. \ref{fig4} and \ref{fig3}. Again for $g_s>0.002$, the distribution is uniform. We studied the cases with $L=150, 400$ and obtained similar results. Our results clearly indicate that for rigid Calabi-Yaus, $\rho(g_s)$ is uniform in the region of interest in Sec. \ref{Sec3}.

From our numerics, we observe that the basic reason for the uniform distribution is the following: As $L$ is increased, generically the number of its divisors increases and as a result the number of points given by \pref{bzf} increases. The first step in the algorithm to bring the points given by \pref{bzf} 
 to the fundamental domain is to act on them repeatedly by $T$ (or $T^{-1}$) so as to bring them to the strip $\hf \leq \tau < \hf$. For large
 $L$, we find that even just after this first step the region of phenomenological interest is uniformly populated with the number of points of the same order as the final answer (i.e the number of points after all points are brought to the fundamental domain). Note that in \pref{bzf}:
$$
{ \rm{Im} }(\tau) = { L \over a_1^2 }\,,
$$
where $a_1$ divides $L$. Thus, for the points given by \pref{bzf}, the number of points with $\rm{Im}(\tau) >1$ is equal to the number of points with $\rm{Im}(\tau)<1$. This is essentially the reason why after the first step in the algorithm the number of points is of the same order as in the final answer. For large $L$, with the increase in the number of divisors, there are more and more points in the region of interest and the spacing between them becomes uniform.

Now, let us turn to the case of general Calabi-Yaus. The exact characterisation of the vacua (the analogue of equation \pref{bzf}) is not available, and a complete numerical analysis remains challenging\footnote{For recent progress in this direction see e.g \cite{Cole:2019enn}} and is beyond the scope of the present work. Here, we will use our results for the case of rigid Calabi-Yaus to develop intuition for the distribution of $g_s$ in case of general Calabi-Yaus (the basic philosophy shall be the same as that advocated in \cite{Bousso:2000xa}). As described in the previous paragraph, the basic reason for the uniform distribution in the case of rigid Calabi-Yaus is that with increase in $L$, generically the number of vacua increases and the solutions are more and more uniformly spaced. This leads to the uniform distribution of $g_s$. For general Calabi-Yaus, the value of the dilaton is set by the ratio of flux quanta associated with the 3-form fluxes $H_3$ and $F_3$. As the number of 3-cycles increases, one can expect the same phenomenon -- the number of vacua increases and the spacing between the values of the dilaton in these solutions decreases and the distribution function for the dilaton becomes uniform. Note that we found a uniform distribution in the case of rigid Calabi-Yaus, where the number of fluxes is only four. For a general Calabi-Yau with large number of 
cycles, the solutions are certainly expected to be more uniformly spaced, corresponding to a uniform distribution of the dilaton.

\section{Soft terms in LVS and KKLT}
\label{AppB}
  
In this section we briefly summarise the structure of the soft masses in KKLT and LVS with matter fields located on D3/D7 branes. The general expressions for the soft masses are given by:
\begin{align}
M_a &= \frac{1}{2}\frac{F^i\partial_if_a}{Re(f_a)}\\
m_{\alpha}^2 &= m^2_{3/2}+V_0-F^{\bar{i}}F^j \partial_{\bar{i}}\partial_{j}\ln(\tilde{K}_{\alpha})\\
A_{\alpha \beta \gamma} &= F^i\left(K_i+\partial_i\ln(Y_{\alpha \beta \gamma})-\partial_i\ln(\tilde{K}_{\alpha}\tilde{K}_{\beta}\tilde{K}_{\gamma}) \right),
\end{align}
where $Y_{\alpha \beta \gamma}$ are the Yukawa couplings, $\tilde{K}_{\alpha}$ is the K\"ahler matter metric and the F-terms are given by $F^i=e^{K/2}K^{i\bar{j}}D_{\bar{j}}\bar{W}$. In the following table we summarise the soft supersymmetry breaking terms in both KKLT and LVS for matter living on D3 and D7 branes \cite{Aparicio:2015psl}.

\begin{table}[h]
\begin{center}
 \begin{tabular}{||c | c | c | c||} 
 \hline
D3 & KKLT & LVS \\ 
 \hline
 $M_{1/2}$ 		& $\frac{3}{2}\frac{1}{a\mathcal{V}^{2/3}}m_{3/2}$  & 	$\frac{3}{4}\frac{\xi}{g_s^{3/2}\mathcal{V}}m_{3/2}$   \\ 
 \hline
 $m_{0}^2$ & $\left(1-3\omega \right)m_{3/2}^2$ &   $\frac{5}{8}\frac{\xi}{g_s^{3/2}\mathcal{V}}m_{3/2}^2$\\
 \hline
 $A_{\alpha \beta \gamma}$ & $-\left(1-s\partial_s\log\left( Y_{\alpha \beta \gamma}\right) \right)\frac{3}{2}\frac{1}{a\mathcal{V}^{2/3}}m_{3/2}$ & $-\left(1-s\partial_s\log\left( Y_{\alpha \beta \gamma}\right) \right)\frac{3}{4}\frac{\xi}{g_s^{3/2}\mathcal{V}}m_{3/2}$ \\
\hline
\hline
D7 & KKLT & LVS \\ [0.5ex] 
 \hline
 $M_{1/2}$ 		& $\frac{1}{a\mathcal{V}^{2/3}} m_{3/2}$  & 	$m_{3/2}$   \\ 
 \hline
 $m_{\alpha}^2$ & $\left(1-3\omega \right)m_{3/2}^2$ &   $\frac{1}{3}m_{3/2}^2$\\
 \hline
 $A_{\alpha \beta \gamma}$ & $-\frac{3}{2}s\partial_s \log\left( Y_{\alpha \beta \gamma}\right)\frac{1}{a\mathcal{V}^{2/3}} m_{3/2}$ & $-m_{3/2}$ \\
  \hline
  \hline
Anomaly & KKLT & LVS \\ [0.5ex] 
 \hline
 $M_{a}$ 		& $-\frac{g_a^2b_a}{16\pi^2}m_{3/2}$  & 	$-\frac{g_a^2b_a}{16\pi^2}M_{1/2}$   \\ 
 \hline
 $m_{i}^2$ & $\sum_a\frac{g_a^4C_a(i)b_a}{(16\pi^2)^2}m_{3/2}^2$ &   $\sum_a\frac{g_a^4C_a(i)b_a}{(16\pi^2)^2}m_{0}^2$\\
 \hline
 $A_{\alpha \beta \gamma}$ & $Y_{\alpha \beta \gamma}\sum_{m= \alpha, \beta, \gamma}\sum_a\frac{C_a(m)}{b_a}M_{a}$ & $Y_{\alpha \beta \gamma}\sum_{m= \alpha, \beta, \gamma}\sum_a\frac{g_a^2C_a(m)}{16\pi^2}A_{\alpha \beta \gamma}$ \\
 \hline
\end{tabular}
\caption{Soft masses for KKLT/LVS with standard model fields realized on D3/D7 branes. Here $g_a$ is the gauge coupling, the parameter $b_a$ is defined as $b_a = 3T_G-T_R$, with the Casimir invariant $T_G$ in the adjoint representation, the Dynkin index $T_R$ and the quadratic Casimir invariants in the fundamental representation $C_a(i)$. The parameter $\omega$ is a function of the coefficients of the K\"ahler matter metrics.}
\end{center}
\end{table}

In the last table we collect the soft masses that come from anomaly mediation. Note that anomaly mediation plays no r\^ole in LVS but is important in KKLT. Let us compute the variation of the soft masses. In the case of KKLT the variation of the gaugino mass scales as:
\be
d M_{1/2} \sim \left(\frac{g_a^2b_a}{16\pi^2}\right)^2\frac{g_s}{\mathfrak{n}^3|\ln W_0|^3} \frac{M_p^2}{M_{1/2}} dN\,,
\ee
which implies a scaling for the number of states:
\be
N_\KKLT (M_{1/2}) \sim \left(\frac{M_{1/2}}{M_p}\right)^2\,.
\ee
This functional dependence holds for all soft masses for D3 and D7 branes. In LVS the gaugino mass at the minimum of the potential scales as:
\be
M_{1/2}=\frac{3}{4}\sqrt{8\pi}\frac{c_1^2}{|W_0|\mathfrak{n}^2}e^{-\frac{2c_2}{g_s\mathfrak{n}}}\,.
\ee 
Performing the variation gives us:
\be
d M_{1/2} \sim \mathfrak{n} M_{1/2}\ln\left(\frac{M_p}{M_{1/2}} \right)^2 d N\,.
\ee
Ignoring subleading logarithmic corrections, this implies a scaling: 
\be
N_\LVS(M_{1/2})\sim \ln\left(\frac{M_{1/2}}{M_p} \right),
\ee
which is again true for all soft masses for D3 and D7 branes. In summary our conclusion for the distribution of the soft terms in KKLT and LVS vacua is:
\be 
\boxed{\,N_\KKLT \sim \left(\frac{M_{\rm soft}}{M_p}  \right)^2\,}\,
\ee
and:
\be 
\boxed{\,N_\LVS \sim \ln\left(\frac{M_{\rm soft}}{M_p}  \right)\,}\,.
\ee

\bibliographystyle{utphys}
\bibliography{mybib}

\end{document}